\documentclass[amsmath,amssymb]{nature}

\usepackage{ifthen}		
\newboolean{SubmittedVersion}
\setboolean{SubmittedVersion}{False} 	

\usepackage{graphics}
\usepackage{graphicx}
\usepackage{amsmath}
\usepackage{amssymb}
\usepackage{color}

\DeclareMathAlphabet{\mathpzc}{OT1}{pzc}{m}{it}

\begin{document}

\title{Observation of a symmetry-protected topological phase with ultracold fermions}

\author{Bo Song$^{\dagger 1}$, Long Zhang$^{\dagger 2,3}$, Chengdong He$^{\dagger 1}$, Ting Fung Jeffrey Poon$^{2,3}$, Elnur Hajiyev$^1$, Shanchao Zhang$^1$, Xiong-Jun Liu$^{*2,3}$ and Gyu-Boong Jo$^{*1}$}

\maketitle

\begin{affiliations}
 \item Department of Physics, The Hong Kong University of Science and Technology, Clear Water Bay, Kowloon, Hong Kong, China
 \item International Center for Quantum Materials, School of Physics, Peking University, Beijing, China
 \item Collaborative Innovation Center of Quantum Matter, Beijing 100871, China
\end{affiliations}


\begin{abstract}
Symmetry plays a fundamental role in understanding complex quantum matters, in particular, in classifying topological quantum phases which have attracted great interests in the recent decade. An outstanding example is the time-reversal invariant topological insulator, a symmetry-protected topological (SPT) phase in symplectic class of the Altland-Zirnbauer classification. Here, we report the first observation for ultracold atoms of a new SPT phase in a one-dimensional optical lattice, and study quench dynamics between topologically distinct phases. The observed SPT phase is protected by a magnetic group and a nonlocal chiral symmetry, with its topology being measured via Bloch states at symmetric momenta. The topology also resides in far-from-equilibrium spin dynamics, which are predicted and observed in experiment to exhibit qualitatively distinct behaviors in quenching to trivial and nontrivial phases, revealing a deep topology-dependent spin relaxation dynamics. This work opens the way to expanding the scope of SPT phases with ultracold atoms and studying non-equilibrium quantum dynamics in such exotic phases.
\end{abstract}

\section*{Introduction}

The discovery of the quantum Hall effect in a two-dimensional (2D) electron gas~\cite{QHE1,QHE2} brought about a new fundamental concept, topological quantum phase whose characterization is beyond Landau symmetry-breaking theory, to condensed-matter physics~\cite{Wen1990}. Recent extensive studies have generated two broad categories of topological matter, the topologically ordered phases~\cite{Kitaev2006,Levin2006,Chen2010} and symmetry-protected topological (SPT) phases\cite{Schnyder2008,Kitaev2009,Ryu2010,Chen2012Science}, which exhibit long-range and short-range quantum entanglement, respectively. Unlike topological orders which are robust against any local perturbations, a SPT phase with a bulk gap has gapless or degenerate boundary modes that are only robust against local perturbations respecting the given symmetries. The earliest examples of SPT phases include the one-dimensional (1D) Su-Schrieffer-Heeger (SSH) model for fermions~\cite{SSH1979} and spin-$1$ antiferromagnetic chain for bosons~\cite{Haldane1983}. The search for new SPT phases has been greatly revived in the recent decade due to the groundbreaking discovery of time-reversal invariant topological insulators in 2D and 3D materials, which exhibit symmetry-protected helical edge or surface modes in the boundary and are characterized by a $\mathbb{Z}_2$ invariant~\cite{Hasan2010,Qi2011}. The topological insulators also inspired generalizations to the SPT phases arising from spatial symmetries, in unconventional superconductors and superfluids~\cite{Chiu2016}.

Despite the broad classes of SPT phases predicted in theory, only a small portion of such phases have been observed in experiment. In comparison with solid state materials, the ultracold atoms are extremely clean systems which can provide controllable platforms in exploring new SPT phases, with many theoretical schemes having been proposed~\cite{Liu2013b,LLu2016,Nonne2013}. To date, some novel topological features of the SSH model have been probed in experiments with ultracold bosons~\cite{Atala:2014,Meier2016}, while the realization of a SPT phase, which has to be in a fermionic system for a noninteracting phase, is not available. Here we propose and realize for the first time a new SPT phase for ultracold fermions which is beyond particle-hole or chiral symmetry protection as usually required in Altland-Zirnbauer (AZ) classification of 1D states~\cite{AZ1997,Chiu2016}.
We  measure the topology via Bloch states at symmetric momenta~\cite{Liu2013a} and the quench dynamics, and unveil a deep topology-dependent mechanism for the spin relaxation in this SPT phase.

\section*{Symmetry-protected topological phases}

\begin{figure}[t]
\begin{center}
\includegraphics[width=170 mm]{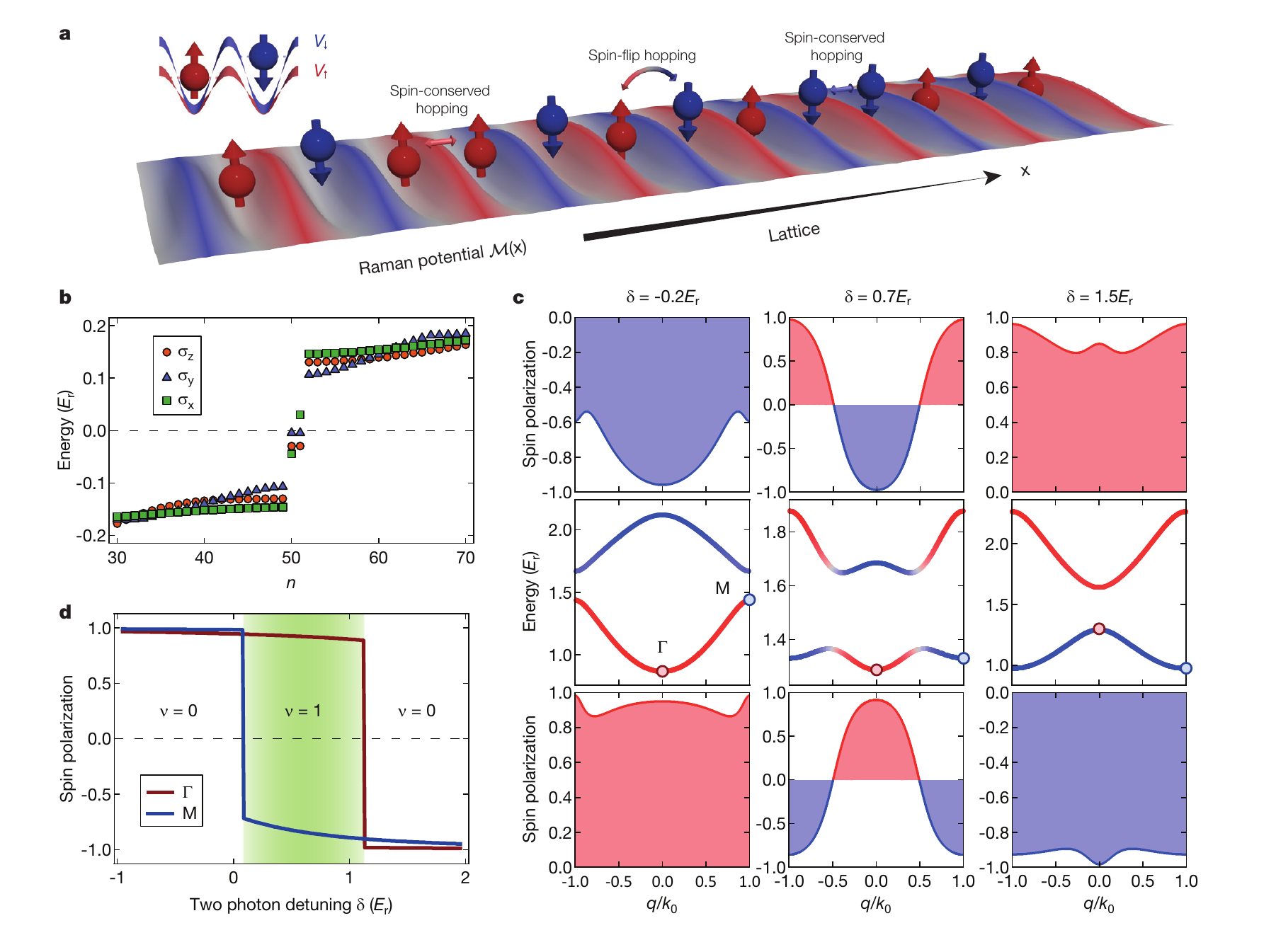}
\caption{ {\bf SPT phases and band topology}.
{\bf a,} Sketch of the 1D Raman lattice model. A spin-dependent lattice potential induces spin-conserved hopping, while the Raman potential contributes to spin-flip hopping. {\bf b,} Bulk and boundary energy spectra obtained by full band diagonalization with a Zeeman perturbation field along the $z$ ($\sigma_z$, red circles), $y$ ($\sigma_y$, blue triangles) and $x$ ($\sigma_x$, green squares) direction, respectively. The in-gap states are boundary modes. The parameters are taken as $V_{\uparrow}=2.5E_{\rm r}$, $V_{\downarrow}=5E_{\rm r}$ and $M_0=1.0E_{\rm r}$ (the same for {\bf c} and {\bf d}).
{\bf c,} Band structure (the second row) and spin textures of both the lowest band (the third row) and the second band (the first row) for $\delta=-0.2E_{\rm r}$ (left), $\delta=0.7E_{\rm r}$ (middle) and $\delta=1.5E_{\rm r}$ (right), respectively.
{\bf d,} The spin polarizations at $\Gamma$ and $M$ points of the lowest band determines the topological trivial ($\nu=0$) and nontrivial ($\nu=1$) regimes.} \label{Fig1}
\end{center}
\end{figure}

We start with the Hamiltonian for spin-$1/2$ ultracold fermions trapped in a 1D optical Raman lattice
potential (Fig.~\ref{Fig1}a), whose realization in experiment will be presented later:
\begin{equation}\label{Ham}
H=\bigr[\frac{p_x^2}{2m}+\frac{V^{\rm latt}_{\uparrow}(x)+V^{\rm latt}_{\downarrow}(x)}{2}\bigr]\otimes{\bf\hat 1}+\bigr[\frac{\delta}{2}+\frac{V^{\rm latt}_{\uparrow}(x)-V^{\rm latt}_{\downarrow}(x)}{2}\bigr]\sigma_z+{\cal M}(x)\sigma_x.
\end{equation}
Here $p^2_x/2m$ is the kinetic energy for motion in $x$-direction,
${\bf\hat 1}$ is a two-by-two unit matrix, $\sigma_{x,y,z}$ are Pauli matrices in spin space, $V^{\rm latt}_{\uparrow,\downarrow}(x)=V_{\uparrow,\downarrow}\cos^2(k_0x)$ denote the 1D optical lattice potentials for spin $|\uparrow,\downarrow\rangle$ states, respectively, with $V_{\uparrow,\downarrow}$
being the lattice depths and $k_0=\pi/a$ ($a$ is the lattice constant), the Raman lattice potential ${\cal M}(x)=M_0\cos(k_0x)$ with amplitude $M_0$, and $\delta$ denotes the two-photon detuning of Raman coupling.

The Hamiltonian (\ref{Ham}) realizes novel topological phases protected by symmetries.
With the lattice potential $V^{\rm latt}_{\uparrow,\downarrow}(x)$ the spin-conserved hopping ($t_{\uparrow,\downarrow}$) is induced, while the Raman potential ${\cal M}(x)$ accounts for the hopping ($t_{\rm so}$) that flips atomic spin (Fig.~\ref{Fig1}a). The Bloch Hamiltonian for the lowest $s$-band physics can be obtained by~\cite{Supplementary} ${\cal H}(\vec m)=-2t_0\cos q_xa\sigma_z+2t_{\rm so}\sin q_xa\sigma_y+\vec m\cdot\sigma+2t_1\cos q_xa{\bf\hat 1}$, where we take $\vec m=(m_x,m_y,m_z)$ for a generic study, $q_x$ is Bloch quasi-momentum, and the hopping coefficients $t_{0/1}=(t_{\uparrow}\pm t_{\downarrow})/2$. It is known that the Hamiltonian ${\cal H}$ with $m_x=0$ exhibits a chiral symmetry defined by $\sigma_x{\cal H}\sigma_x=-{\cal H}$ in the spin-independent lattice regime $V_{\uparrow}^{\rm latt}=V_{\downarrow}^{\rm latt}$, which gives $t_{\uparrow}=t_{\downarrow}$, and belongs to the AIII class according to the AZ classification~\cite{AZ1997,Liu2013b}. The topology of such AIII class phase is quantified by an integer winding number ($\mathbb{Z}$) in the single particle regime, pictorially characterized by the circles that spin of Bloch states winds over the first brillouin zone (FBZ), and determines the number of midgap end states due to bulk-boundary correspondence. Furthermore, in the interacting regime, it was shown that the topology of the 1D AIII class topological insulator is reduced from $\mathbb{Z}$ to $\mathbb{Z}_4$~\cite{Liu2013b}, which is closely related to the Fidkowski-Kitaev $\mathbb{Z}_8$-classification of 1D interacting Majorana chains of BDI class~\cite{Fidkowski2011}. The AIII class topological insulator has not been studied previously in experiment.

The striking result is that a new topological phase can be achieved beyond the chiral symmetry protection when the lattice $V^{\rm latt}_{\uparrow,\downarrow}$ and $t_{\uparrow,\downarrow}$ are spin-dependent, in which regime no chiral symmetry preserves and the bulk energy spectrum is asymmetric. From the numerical results in Fig.~\ref{Fig1}b one finds that
the midgap and degenerate end states are obtained for $\vec m=m_y\hat e_y$ and $\vec m=m_z\hat e_z$, respectively, while only a nonzero $m_x$ term splits out the degeneracy. The existence of midgap or degenerate end states manifests new SPT phases, as shown below. We consider first the case with $\vec m=0$. One can verify that ${\cal H}(\vec m=0)$ satisfies a magnetic group symmetry defined as the product of time-reversal and mirror symmetries~\cite{Supplementary}, giving $M_x=\sigma_zK\otimes R_x$ with $M_x{\cal H}(q_x)M_x^{-1}={\cal H}(-q_x)$, where $K$ is complex conjugate and $R_x$ denotes the spatial reflection along $x$ axis. Besides, the Hamiltonian also satisfies a {\it nonlocal chiral symmetry} defined as ${\cal S}=\sigma_z\otimes T_x(k_0)\otimes R_x$, where $T_x(k_0)$ is a $k_0$-momentum translation, giving that ${\cal S}{\cal H}(q_x){\cal S}^{-1}=-{\cal H}(q_x)$. The midgap end states at left ($|\psi_{L}\rangle$) and right ($|\psi_R\rangle$) hand boundaries are protected by the both symmetries under a novel mechanism. First, by definition both $M_x$ and ${\cal S}$ transform one of $|\psi_{L/R}\rangle$ to the other. Moreover, their commutation and anti-commutation relations with ${\cal H}$ imply that the magnetic group (nonlocal chiral) symmetry connects two states with identical (opposite) energy. For this $|\psi_{L/R}\rangle$ have to be zero energy, i.e. midgap states in the presence of both symmetries, which is the case for $\vec m$ along $y$ axis. The $m_z$ term keeps $M_x$ symmetry, hence protects the end state degeneracy, while $m_x$ term breaks $M_x$ symmetry and splits out the degeneracy, leading to the end state spectra in Fig.~\ref{Fig1}b. In the experiment a nonzero $m_z$ term can be easily engineered by manipulating the two-photon detuning $\delta$ of the Raman coupling. In this regime the magnetic group ensures that Bloch states at the two symmetric momenta $\{\Lambda_j\}=\{\Gamma(q_x=0),M(q_x=\pi/a)\}$ are eigenstates of $\sigma_z$, having spin polarization $P(\Lambda_j)=\pm1$, with which a $\mathbb{Z}_2$ invariant~\cite{Liu2013a} $(-1)^\nu=\Pi_j{\rm sgn}[P(\Lambda_j)]$ can be introduced to characterize the band topology.
The topologically nontrivial (trivial) phase corresponds to $\nu=1$ (0). Fig.~\ref{Fig1}c shows the spin texture with different parameter conditions, from which the $\mathbb{Z}_2$ invariant can be read out directly, and Fig.~\ref{Fig1}d presents the phase diagram versus $\delta$.

\section*{Quench dynamics}

We examine the spin dynamics after a quench from one SPT phase to another. The quench is performed by ramping suddenly the
two-photon detuning from $\delta_i$ to $\delta_f$, with $\delta_i$ and $\delta_f$ corresponding to two topologically distinct regimes.
The quantum dynamics is captured by the time-dependent density matrix $\rho(t)$,
which satisfies the Lindblad master equation~\cite{Lindblad,Hu2016}
\begin{equation}\label{Lin_eqn}
\dot{\rho}=-\frac{i}{\hbar}[H',\rho]+\gamma\left(L\rho L^\dagger-\frac{1}{2}\{L^\dagger L,\rho\}\right),
\end{equation}
where $\gamma$ denotes the noise-induced decay rate and $L$ is the Lindblad operator characterizing
effects of the environment. For a real ultracold atom system, we consider an external trapping potential for the study and the Hamiltonian
$H'=H+V_{\rm trap}$, with $V_{\rm trap}(x)=\frac{1}{2}m\omega_x^2x^2$ and $\omega_x$ being the trapping frequency.
The sharp contrast between topological and trivial phases is that the spin of Bloch states over FBZ winds over all direction in $y$-$z$ plane in the former case, while is polarized to $+z$ (or $-z$) direction in the latter (Fig.~\ref{Fig1}c). This feature accounts for different fundamental mechanisms governing the quench spin dynamics, as given below.

\begin{figure}[t]
\begin{center}
\includegraphics[width= 120 mm]{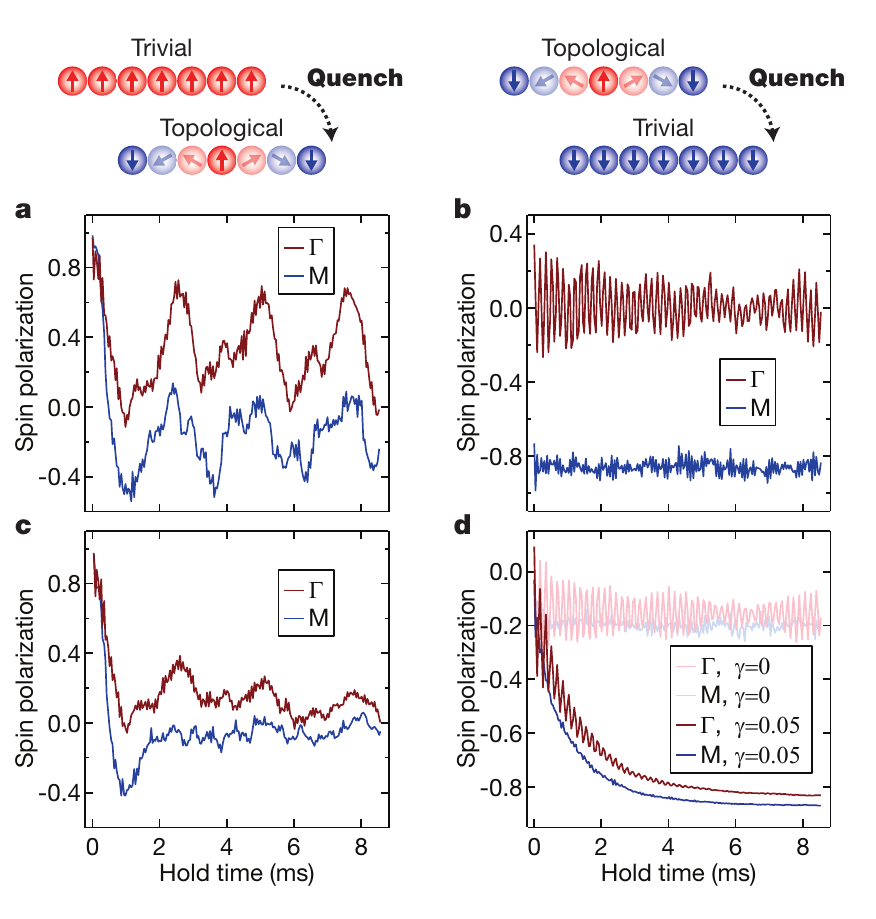}
\end{center}
\caption{{\bf Quench dynamics of SPT phases}. The numerical simulation is performed for quench process from trivial to topological regimes with $\delta_i=-0.2E_{\rm r}$ and $\delta_f=0.7E_{\rm r}$ in ({\bf a,c}), and from topological to trivial regimes with $\delta_i=0.7E_{\rm r}$ and
$\delta_f=2.0E_{\rm r}$ in ({\bf b,d}). In all the cases the trapping frequency reads $\omega_x=(2\pi)300$Hz, the optical Raman lattice potentials $V_{\uparrow}=2.5E_{\rm r}$, $V_{\downarrow}=5E_{\rm r}$, and $M_0=1.0E_{\rm r}$.
{\bf a-b,} Quench dynamics for spin polarizations at $\Gamma$ and $M$ points at zero temperature and without dissipation.
In {\bf c}, quench spin dynamics from trivial to topological regime at the temperature $T=50$nK, with the noise-induced dephasing rate
$\gamma=0.005$;
in {\bf d}, quench spin dynamics from topological to trivial at $T=130$nK,
with dissipation rate taken as $\gamma=0$ (dash-dotted lines) and $\gamma=0.05$ (solid lines), respectively.}
\label{Fig2}
\end{figure}

We consider first the case without dissipation ($\gamma=0$). In this regime the time-evolution is unitary, and the spin dynamics are governed by the following two quantum processes, namely, the interband transition at each quasi momentum $q_x$ due to nonequilibrium population after quench, and the intraband transition induced by trapping potential between Bloch states of different momenta.
The numerical simulation of the spin dynamics at $\Gamma$ and $M$ points are shown in Fig.~\ref{Fig2}a and Fig.~\ref{Fig2}b for quench to topological ($\delta_f=0.7E_{\rm r}$) and trivial ($\delta_f=2.0E_{\rm r}$) phases, respectively. In both cases, the spin dynamics exhibit a fast oscillation due to interband transitions, with the oscillation frequency determined by the local band gap at $\Gamma$ and $M$ points. However, the spin dynamics for quench to topological phase exhibit also a slower spin-wave like collective oscillation associated with a fast decay in the very beginning stage, which is absent in the quench to trivial phase. This novel behavior is a consequence of the topological spin texture or spin-orbit fields in momentum space~\cite{Supplementary}. In the topological regime, the intraband transitions induced by trapping potential randomize the fast oscillation caused by interband transitions due to the strongly momentum-dependent spin-orbit fields, leading to a fast decay in the very beginning and leaving only a collective oscillation governed by the trapping potential. Instead, in the trivial regime, all the spin states and spin-orbit fields of FBZ approximately point to the $+z$ (or $-z$) direction. Thus the trapping-induced intraband transitions have negligible effect on the spin oscillation.

We further show in Fig.~\ref{Fig2}c,d the quench dynamics with dissipation, as captured by the non-unitary part of Eq.~(\ref{Lin_eqn}).
For quench to the trivial phase, since the spin dynamics are dominated by the interband oscillations, the dissipation mainly results from the decay of the excited bands. In comparison, for quench to the topological phase, the dissipation mainly describes a spin dephasing of the collective oscillation induced by trapping. The Lindblad operator $L$ then takes the form $L=(\widetilde{\sigma}_x+i\widetilde{\sigma}_y)/2$ (for decay) and $L=\widetilde{\sigma}_z$ (for dephasing), respectively, and the latter effect is typically much weaker than the former~\cite{Hu2016}. Here $\widetilde{\sigma}_{x,y,z}$ denote the
Pauli matrices in the eigenbasis of the post-quench Hamiltonian $H(\delta_f)+V_{\rm trap}$. The numerical simulation shows that the quench dynamics from trivial to topological regime resembles the one without dissipation except for a small decay in the amplitude of collective oscillation, followed by a quick interband balance in the very beginning (Fig.~\ref{Fig2}c), while the dynamics in the other way round renders a pure decay from initial state to the final equilibrium phase (Fig.~\ref{Fig2}d). With these results we draw a novel conclusion that the dissipation and trapping potential have dominant (weak) effects on the spin dynamics with quench to trivial (topological) and topological (trivial) phases, respectively, due to the topological and trivial spin textures of the final SPT phase, as observed in the present experiment.

\section*{Experimental Result}

\paragraph*{\textbf{Optical Raman lattices for ytterbium atoms}}

\begin{figure}[t]
\begin{center}
\includegraphics[width= 110 mm]{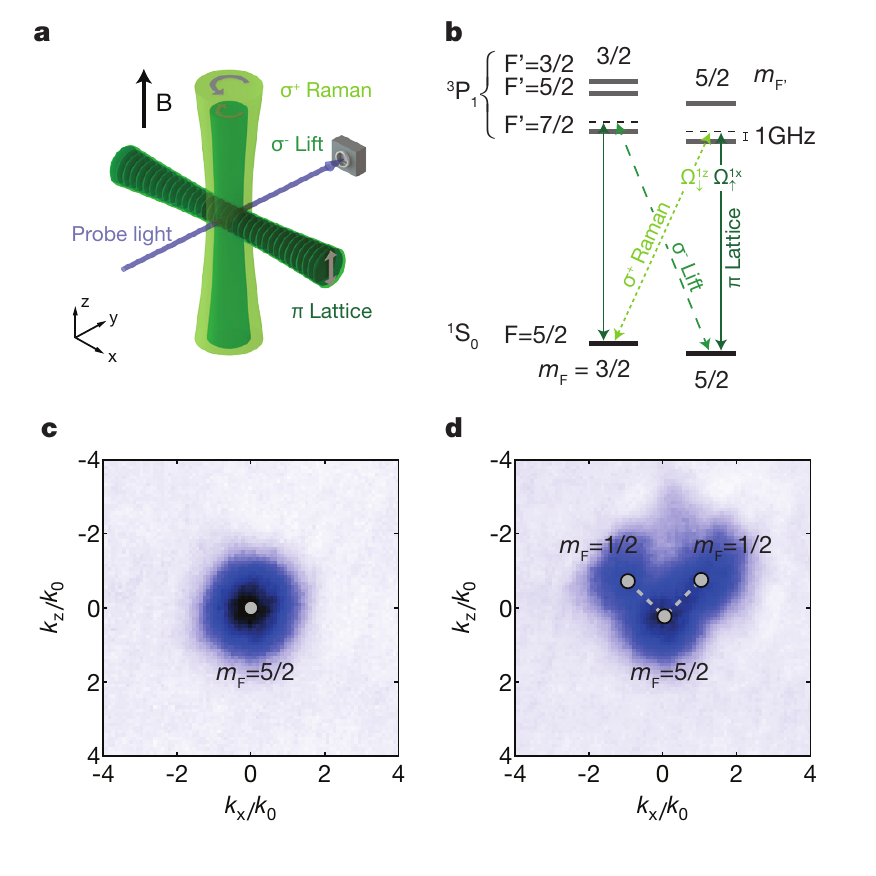}
\end{center}
\caption{ \textbf{A sketch of the experimental set-up}
\textbf{a}, The experimental set-up consists of a one-dimensional optical lattice with lattice along $x$ direction and a perpendicular Raman beam generating a periodic Raman potential $\mathcal{M}(x)=M_0\cos(k_0 x)$. A circularly-polarized beam is added along the quantized axis in $z$ direction introducing the spin-dependent AC Stark shift.  \textbf{b}, Both the lattice and the Raman beams are near blue-detuned from the $F=\frac{5}{2} \to F'=\frac{7}{2}$ inter-combination transition, and induce the Raman transition between  $|\uparrow\rangle=|\frac{5}{2},\frac{5}{2}\rangle$ and $|\downarrow\rangle=|\frac{5}{2},\frac{3}{2}\rangle$ hyperfine states of the $^1$S$_0$ ground manifold.  \textbf{c}, Time-of-flight image of $m_F=5/2$ atoms without pulsing optical Raman lattice beams. The direction of the gravity is along $z$ direction. \textbf{d}, When the optical Raman lattice is briefly switched on in such a way that $m_F=1/2$ atoms are coupled out, two $m_F=1/2$ clouds are relatively shifted by $\sim$2$k_0$ along $x$ direction.}
\label{Fig3}
\end{figure}

To realize the Hamiltonian (\ref{Ham}) in experiment, we utilize a 1D optical lattice dressed by a periodic Raman coupling potential, making up a so-called
optical Raman lattice~\cite{Liu2013b,Wu2016}. The optical Raman lattice is generated by the use of the intercombination $\lambda_0=2\pi/k_0=$ 556~nm transition of $^{173}$Yb atoms, blue-detuned by $\sim$1~GHz from the $^1$S$_0 (F = \frac{5}{2})\leftrightarrow {}^3$P$_1 (F' = \frac{7}{2})$ transition. The lattice potential produced by counter-propagating lights, being linearly-polarized along the $z$ direction, forms a spin-dependent potential $V_{\sigma}^{\rm latt}(x)=V_{\sigma}\cos^2(k_0x)$,
where $V_{\sigma=\{\uparrow,\downarrow \}}=\sum_{F'}\hbar\frac{|\Omega^{1x}_{\sigma,F'}|^2}{4\Delta_{F'}}$ with $|\uparrow\rangle=|m_F=\frac{5}{2}\rangle$ and $|\downarrow\rangle=|m_F=\frac{3}{2}\rangle$~\cite{Song:2016ep}.  The effective Rabi frequencies $\Omega^{1x}_{\sigma,F'}$ and the single-photon
detunings $\Delta_{F'}$ are determined from all relevant transitions to the excited $F'=(\frac{7}{2},\frac{5}{2},\frac{3}{2})$ states in
the $^3$P$_1$ manifold~\cite{Supplementary}.  With the quantized axis  set along the $z$ direction, a standing-wave lattice light of frequency $\omega_0=k_0 c$ ($c$ is the speed of light), denoted as $ \textbf{E}_{1x} \propto \textbf{e}_{z} 2\overline{E}_{1x} e^{-i\omega_0t}\text{cos}(k_{0}x) $ where $\overline{E}_{1x}$ is the amplitude of the light field and $\textbf{e}_z$ is the unit vector in $z$-direction,  induces a
$\pi$-transition between the ground and the excited states, thus resulting in the relation $V_{\downarrow}/V_{\uparrow}=3.84$.

Another circularly polarized light of frequency $\omega_1$, denoted as $ \textbf{E}_{1z} \propto (\textbf{e}_{x} + i\textbf{e}_{y})\overline{E}_{1z} e^{i(k_{0}z-\omega_1t)}$,
is applied along the $z$ direction, and induces the Raman coupling between $|\uparrow\rangle$ and $|\downarrow\rangle$ states with the Rabi frequency $\Omega^{1z}_{\downarrow,F'}$. Both the lattice and Raman lights are generated by a single laser source and controlled by
phase-locked acousto-optical modulators (AOMs), making the frequency difference $\delta \omega=\omega_0-\omega_1$ tunable.
The Raman coupling potential $\mathcal{M}(x,z)$ takes the form
$\mathcal{M}(x,z)\propto M_0 e^{ik_0 z}(e^{ik_0 x}+e^{-ik_0 x})$, where $M_0=\sum_{F'}\Omega^{1x}_{\uparrow,F'}\Omega^{1z}_{\downarrow,F'}/2\Delta_{F'}$. Since we focus on the motion along the $x$ direction, which is
uncoupled to the other two directions, we neglect the irrelevant phase $e^{ik_0 z}$ and write the Raman potential as
$\mathcal{M}(x)=M_0 \cos(k_0x)$. To separate out spin-$1/2$ space from other hyperfine states of the ground manifold, an additional circularly-polarized 556~nm  light, $ \textbf{E}_{2z} \propto (\textbf{e}_{x} - i\textbf{e}_{y})\overline{E}_{2z} e^{ik_{0}z}$ (called as a lift beam), is applied along the $z$ direction to induce spin-dependent AC Stark shift~\cite{Song:2016ep}. Note that the spin-dependent lattice potential $V_{\uparrow,\downarrow}^{\rm latt}(x)$ adds an additional energy shift to the Bloch bands corresponding to two spin states, which can be readily compensated by the two-photon detuning. Finally a  Zeeman term  is given as $m_z=({\delta-\delta_o})/{2}$, where $\delta_0$ is the on-site energy difference in the spin-dependent lattice~\cite{Supplementary}.



\paragraph*{\textbf{Symmetry in an optical Raman lattice }}
To investigate a new SPT phase in experiment, it is crucial to test that the system has a mirror symmetric optical Raman lattice and a Zeeman term $\vec m$ is parallel to the mirror plane, leading to a magnetic group symmetry. In the experiment, the two-photon detuning $\delta$ sets a nonzero $m_z$, but a finite $m_x$ term  may be generated when the Raman potential $\mathcal{M}(x)$ breaks a mirror symmetry of the lattice and thus induces on-site spin-flipping (see a supplementary note). When the Raman beam $\textbf{E}_{1z}$ is not perfectly along the $z$ direction with the wave vector $\hat{k}=k_0 \textbf{e}_z+k_{\perp,x}\textbf{e}_x+k_{\perp,y}\textbf{e}_y$, the Raman potential has an additional phase factor $\mathcal{M}(x)=M_0 \cos(k_0x)e^{ik_{\perp,x}x}$, therefore breaking a mirror symmetry unless $k_{\perp,x}$ is close to zero.

We begin with spin-polarized $m_F=5/2$ fermions followed by a brief pulse of the optical Raman lattice beams, and then examine the wave number $k_{\perp,x}$ of the Raman beam $\textbf{E}_{1z}$.  After the time-of-flight expansion, $m_F=5/2$ atoms are shifted downward (in $-\hat{z}$ direction) while $m_F=3/2$ atoms upward (in $+\hat{z}$ direction) which manifests spin-orbit coupling induced by the Raman transition~\cite{Lin:2011hn,Wang:2012gv,Cheuk:2012id,Burdick:2016jt,Song:2016ep,Mancini2015,Li2016}. Along the $x$ direction, however, $m_F=3/2$ atoms absorb the momentum shift of $\pm 1/2 \hbar k_0$ resulting two small $m_F=3/2$ clouds after time-of-flight.  In the experiment, we spin-orbit-couple multiple hyperfine states $m_F=5/2,3/2,1/2$ in such a way that two small daughter clouds of $m_F=1/2$ atoms are well separated from the $m_F=5/2$ atoms. Finally we assess the left-right symmetry of the atom diffraction and determine the angle $\theta$ between the lattice and the Raman beam to be $\theta=$89.3$\pm$1.5 or the wavenumber $k_{\perp,x}=0.01k_0$, which ensures negligible $m_x$ term in the Hamiltonian (\ref{Ham}) and therefore a magnetic group symmetry.

\begin{figure}[t]
\begin{center}
\includegraphics[width= 175 mm]{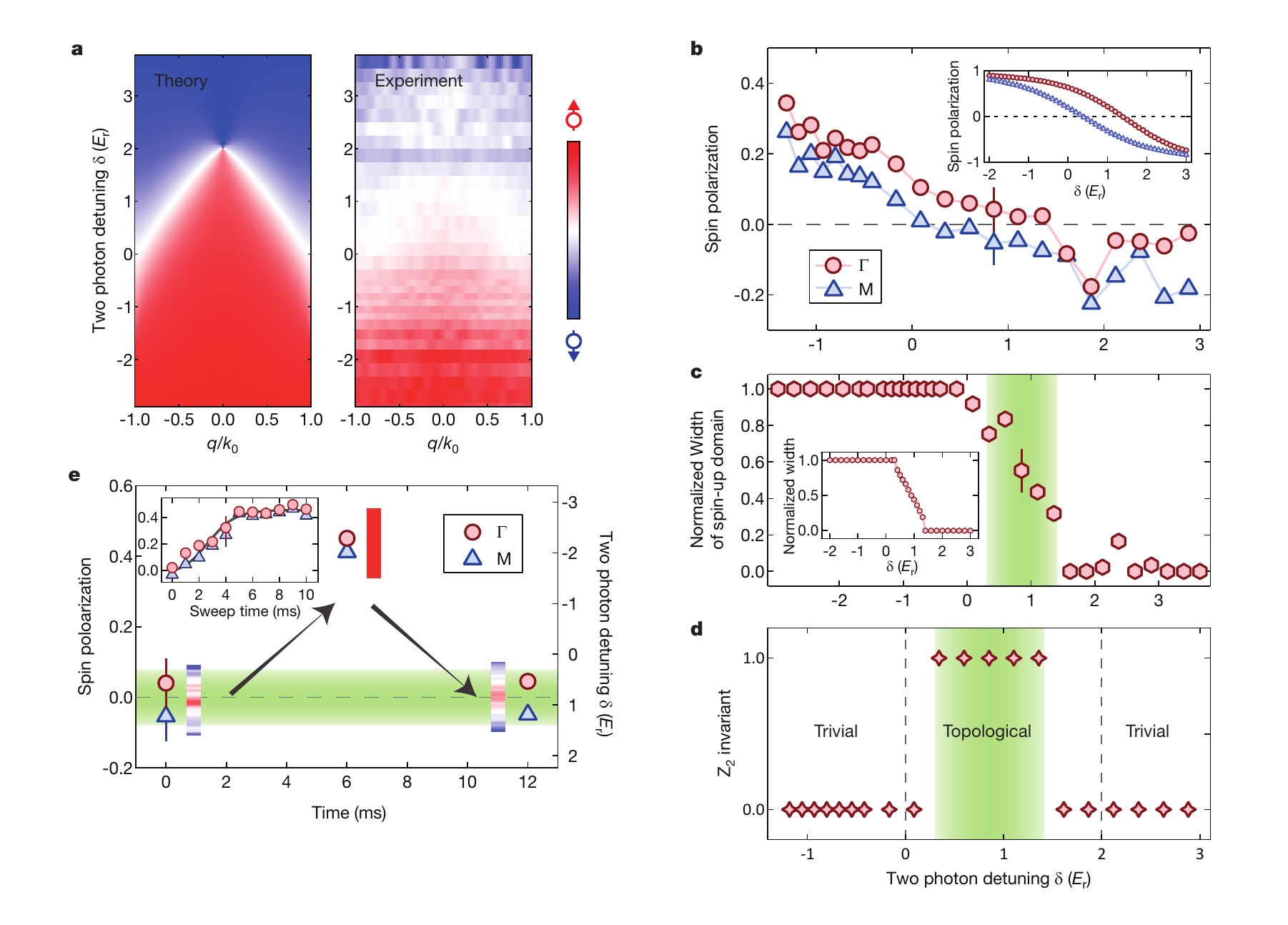}
\end{center}
\caption{\textbf{ In-equilbirum spin textures and their $\mathbb{Z}_2$ invariant across the topology phase transition}
\textbf{a}, Experimental measurement of spin polarization $P(q_x)=(n_{\uparrow}(q_x)-n_{\downarrow}(q_x))/(n_{\uparrow}(q_x)+n_{\downarrow}(q_x))$ within FBZ as a function of two-photon detuning $\delta$. For all measurements, the lattice depth and the Raman coupling strength are set to $V_{\uparrow}$=1.1(1)$E_r$, $V_{\downarrow}$=4.2(4)$E_r$ and  $M_0$=1.88(10)$E_r$. Numerical calculation at $T=0$~nK is shown for comparison.  \textbf{b}, The spin polarization is obtained at the symmetry points $\Gamma$ and $M$. The inset figure gives numerical calculation based on experimental parameters. \textbf{c}, Coexistence of $|\uparrow\rangle$ and $|\downarrow\rangle$ within the FBZ is visualized by measuring the width of the $|\uparrow\rangle$ domain of the spin texture normalized by 2$\hbar k_0$ as a function of $m_z$.  \textbf{d}, The measured $\mathbb{Z}_2$ invariant $\nu$  reveals the SPT phase, which is consistent with the predicted region (shaded region) for the current experiment parameters.   The vertical dashed line indicates the predicted phase transition point at $T=0$~nK. \textbf{e}, A SPT topological phase, adiabatically prepared at $\delta$=0.8(3)$E_r$ initially, is converted into a trivial phase at $\delta$=-2.1(3)$E_r$, and restored back to the original SPT phase. The dashed line of the inset is a guide for eyes. }
\label{Fig4}
\end{figure}

\paragraph*{\textbf{Observation of the symmetry-protected topological phase}}
We now observe a SPT phase for ultracold fermions by loading a equal mixture of $|\uparrow\rangle$ and $|\downarrow\rangle$ atoms, prepared at $T/T_F = $~0.4 in a {crossed optical dipole trap (ODT) where $T_F$ is the Fermi temperature of the system~\cite{Song:2016ep}, into the optical Raman lattice. An optical AC Stark shift, separating out an effective spin-1/2 subspace from other hyperfine levels, is applied within 5~ms before the optical lattice potential and the Raman-dressing beams are adiabatically switched on with an 10~ms exponential ramp to the final value. During the ramp, the crossed ODT power is  increased to compensate the anti-trapping effect arising from the blue-detuned Raman potential. Consequently the temperature is increased to $T=$~125(20)~nK when all optical potentials with $V_{\uparrow}$($V_{\downarrow}$)=1.1 $E_r$ (4.2 $E_r$) and Raman coupling $M_0$=1.88$E_r$, where $E_r=\hbar^2 k_0^2/2m$ and $m$ is the mass of ytterbium atom, are turned on.  In the experiment, the spin-resolved time-of-flight (TOF) imaging is taken after all the laser beams are suddenly switched off, followed by spin-sensitive blast lights~\cite{Song:2016ep}.

The crucial feature of SPT phases is the non-trivial spin texture within the FBZ.   Fig.~\ref{Fig4}a shows measured spin textures $P(q_x)=(n_{\uparrow}(q_x)-n_{\downarrow}(q_x))/(n_{\uparrow}(q_x)+n_{\downarrow}(q_x))$ within the FBZ  reconstructed from the integrated momentum distribution along the $\hat{x}$-directional for different values of the two-photon detuning.  As the two-photon detuning $\delta$ is scanned from the negative to positive value, the overall spin polarization of the lowest $s$-band gradually changes from $|\uparrow\rangle$ (red) to $|\downarrow\rangle$ (blue). However when the Raman transition resonantly couples the two lowest $s$-band corresponding to $|\uparrow\rangle$ and $|\downarrow\rangle$ respectively, the spin polarization changes from one to the other within the FBZ as shown in Fig.~\ref{Fig4}b.  Finally we use the connection between the $\mathbb{Z}_2$ invariant and the $P(\Gamma, M)$, and determine $ (-1)^{\nu}=\prod_{i=M,\Gamma} sgn[P(\Lambda_i)]$  by measuring averaged spin polarization  $P(\Lambda_i)=\frac{10}{k_r}\int_{\Lambda_i-0.05k_r}^{\Lambda_i+0.05k_r}P(q')dq'$  that takes into account the optical resolution of the imaging system~\cite{Liu2013a}. Fig.~\ref{Fig4}d shows the emergence of SPT phases ($\nu=1$) out of trivial phases ($\nu=0$). The spin-dependent lattice shifts the topological regime due to the on-site energy difference, and furthermore the finite temperature effect reduces the topological regime in good agreement with the prediction (indicated by shaded region in Fig.~\ref{Fig4}d).


\paragraph*{\textbf{Adiabatic control of the band topology}}
In a second set of experiment, we highlight the adiabatic preparation of the topological phase starting from the trivial regime and vice versa. In the experiment, a topological phase is adiabatically prepared at $\delta=0.8(3)E_r$  as described in the previous experiment and subsequently the two-photon detuning $\delta$ is linearly ramped to the final value of $\delta=-2.1(3)E_r$ where the trivial phase is expected in equilibrium (Fig.~\ref{Fig4}a).  Here we test the adiabaticity of the sweep process by probing the spin polarization for different sweep times as shown in the inset of Fig.~\ref{Fig4}e. The value of the spin polarization at symmetric points is identical to the equilibrium case when the sweep time is longer 5~ms. As a final examination of the adiabatic control, we show that the SPT phase can be converted into the trivial phase, and restored back to the topological one as in Fig.~\ref{Fig4}e. The spin texture and its invariant measured at each time (0, 6, 12)~ms is consistent with the equilibrium case.

\begin{figure}[t]
\begin{center}
\includegraphics[width= 120 mm]{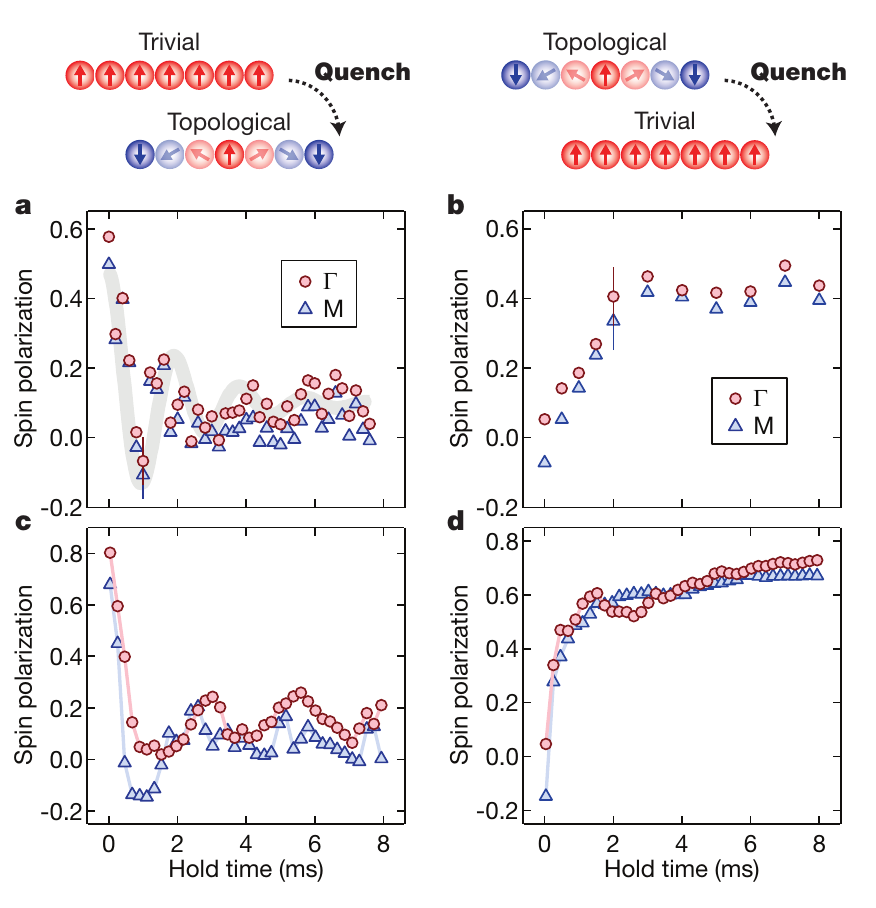}
\end{center}
\caption{\textbf{Far-from-equilibrium spin dynamics after  quench to SPT and trivial phases}
 {\bf a-b,} The measured spin dynamics after a quench from trivial to topological (a) or from topological to trivial (b).
The parameters are $V_{\uparrow}=1.1(1)E_r$, $V_{\downarrow}=4.2(4)E_r$ and $M_0=1.88(10)E_r$. The two-photon detuning is set to be $\delta=-2.1(3)E_{\rm r}$ and $0.8(3)E_{\rm r}$ for the trivial and topological regime, respectively. The shaded region serves as a guide for eyes. {\bf c-d,}  Numerical calculations of the quench dynamics. Parameters all take
the same values as in (a-b) except $\delta=-1.5E_{\rm r}$ ($\delta=0.8E_{\rm r}$) for the trivial (topological) regime. The temperature is set at $T=150$nK. }
\label{Fig5}
\end{figure}

\section*{Topology-dependent spin relaxation after quench }
Having known spin textures in equilibrium and their topological invariants, we now turn our attention to  the far-from-equilibrium spin dynamics after the quench between the trivial and the topological regime.  Using dynamic control of the Hamiltonian, we in particular consider a bi-directional quench between the topological regime ($\delta=$0.8(3)$E_r$) and spin-$|\uparrow\rangle$ trivial regime ($\delta=-$2.1(3)$E_r$) with the trap frequency $\omega_x=2\pi\times 241(20)$~Hz, and moitor the spin dynamics under the post-quench Hamiltonian. Fig.~\ref{Fig5} shows experimental measurements of spin polarization at symmetric points after bi-directional quench, which indeed reflects the topological nature of the post-quench Hamiltonian.  If the system is quenched to the SPT phase with spin-orbit field (Fig.~\ref{Fig5}a), the $q_x$-dependent effective magnetic (spin-orbit) field leads to spin-wave-like collective dynamics with spin relaxation to the final state. This is in sharp contrast to the case that the system is quenched to the trivial phase where the spin is averagely polarized to z direction, and the dominant spin dynamics shows monotonous relaxation (Fig.~\ref{Fig5}b). It can be seen that the collective oscillation exhibits a frequency larger than the trapping frequency. This is because the trapping-induced intraband transition can have an energy detuning between Bloch states at different momenta, which enhances the collective oscillation frequency~\cite{Supplementary}. The observed topology-dependent spin dynamics shows a quite good agreement with numerical calculations taking into accounts the trap geometry, the finite temperature effect and the dissipation process (Fig.~\ref{Fig5}c,d). The spin relaxation dynamics clearly reflect the topological property of the phase after quench.


\section*{Discussion and Conclusion}

The novel SPT phase we uncovered here highlights the great capabilities to realize and explore new topological states with ultracold atoms. In particular, the study may open the way to observing all SPT phases of 1D in the AZ classification~\cite{Ryu2010}, including the chiral topological phase of AIII class which can be readily achieved by considering spin-independent rather than spin-dependent optical lattice in Eq.~\eqref{Ham}~\cite{Liu2013b}, the BDI class and D classes of superfluids by adding attractive interactions~\cite{He2014}, and also the exotic phases protected by nonsymmorphic symmetries~\cite{Chen2016,zhang2016}. Compared with the solid state materials whose complicated environment causes difficulty in general in engineering the symmetries, the full steerability of ultracold atoms can enable a precise study of broad classes of SPT phases, of which a most fascinating issue is to probe the reduction of topological classification of SPT phases from single particle to interacting regimes and has been extensively discussed in theory~\cite{Ryu2010,Fidkowski2011,Gu2014,Morimoto2015}. Such important issue is very difficult to be investigated in solid state experiments due to the great challenges of realization, but is promising for ultracold atoms based on the current study with properly engineered interactions~\cite{Liu2013b,Zhou2016}.


Further generalization of the present study to higher dimensional systems also offers the simulation of quantum phases beyond natural conditions. For example, 2D tunable Dirac semimetals driven by spin-orbit coupling can be readily achieved by applying the present scheme to 2D systems~\cite{Poon2017}, and are of particular interests since a spin-orbit coupled 2D Dirac semimetal is still not available in solid state materials~\cite{Young2015}. By tuning attractive interactions the novel topological Fulde-Ferrell superfluid~\cite{Qu2013,Zhang2013} and Majorana zero modes protected by Chern-Simons invariant~\cite{Chan2017} have been predicted in such 2D Dirac semimetals. Finally, the nonequilibrium quench dynamics unveiled in this work show novel topology-dependent spin relaxation dynamics. Generalization of the quench study to higher dimensional systems is expected to bring about more nontrivial quantum spin dynamics~\cite{Wang2017}, which may provide insights into the understanding of nonequilibrium many-particle dynamics in the exotic phases.

\section*{References}

\bibliographystyle{naturemag}

\bigskip

\begin{addendum}
 \item {We thank T.-L. Ho, Z. Qi, K.T. Law, S.Z. Zhang and C.-K Chan for useful discussion. We also thank Yueyang Zou, Wei Huang and Geyue Cai for experimental assistance. G.-B. J. acknowledges the generous support from the
Hong Kong Research Grants Council and the Croucher
Foundation through ECS26300014, GRF16300215,
GRF16311516 and the Croucher Innovation grants
respectively. X.-J.L acknowledges the support from MOST (Grant No. 2016YFA0301604), NSFC (No.
11574008), and Thousand-Young-Talent Program of China.}
 \item[Competing Interests] The authors declare that they have no
competing financial interests.
 \item[Author Contributions] $^{\dagger}~~$B.S, L.Z and C.H have contributed equally to this work.
 \item[Author Information] Correspondence and requests for materials should be addressed to  X.-J. Liu (xiongjunliu@pku.edu.cn) or G.-B. Jo (gbjo@ust.hk).
 \end{addendum}

\ifthenelse{\boolean{SubmittedVersion}}{\processdelayedfloats}{\cleardoublepage}

\pagebreak
\clearpage
\begin{center}
\textbf{\Large Supplementary Materials}
\end{center}

\makeatletter
\setcounter{equation}{0} \setcounter{figure}{0}
\setcounter{table}{0} \setcounter{page}{1} \makeatletter
\renewcommand{\theequation}{S\arabic{equation}}

\small
\openup -0.8em

\section{Energy spectrum and spin polarization}

The one-dimensional (1D) optical Raman lattice Hamiltonian $H$ [Eq.~(1) in the main text] can be exactly
diagonalized in the plane-wave basis set $\{\Phi_{\alpha\uparrow}(q_x),\Phi_{\beta\downarrow}(q_x)\}$ with
\begin{equation}
\Phi_{\alpha\uparrow}(q_x)=\frac{1}{\sqrt{L}}e^{i(q+2k_0\alpha)x},\quad\Phi_{\beta\downarrow}(q_x)=\frac{1}{\sqrt{L}}e^{i(q+2k_0\beta+k_0)x}.
\end{equation}
Here, $\alpha$ and $\beta$ are both integers, and $L$ denotes the lattice length.
Then the eigenfunctions for the Hamiltonian with eigenvalue $E_n(q_x)$ can be expressed as
\begin{equation}
|\Psi_n(q_x)\rangle=\sum_{\alpha}A^{(n)}_{\alpha}\Phi_{\alpha\uparrow}(q_x)|\uparrow\rangle+\sum_{\beta}B^{(n)}_{\beta}\Phi_{\beta\downarrow}(q_x)|\downarrow\rangle,
\end{equation}
with $A^{(n)}_{\alpha}$ and $B^{(n)}_{\beta}$ the coefficients for the $n$th band. Using the relation $\langle\Phi_{\alpha'\sigma'}|\Phi_{\alpha\sigma}\rangle
=\delta_{\alpha',\alpha}\delta_{\sigma',\sigma}$ ($\sigma=\uparrow,\downarrow$), one can write the Hamiltonian in the matrix form and
diagonalize $H$ numerically. With the Bloch eigenstates obtained by exact diagonalization, the spin polarization of the $n$th band at Bloch momentum $q_x$ can be obtained by
\begin{eqnarray}
\bar \sigma_z^{(n)}(q_x)&=&\langle\Psi_n(q_x)|\sigma_z|\Psi_n(q_x)\rangle=\sum_{\alpha}|A^{(n)}_{\alpha}\Phi_{\alpha\uparrow}(q_x)|^2-\sum_{\beta}|B^{(n)}_{\beta}\Phi_{\beta\downarrow}(q_x)|^2.
\end{eqnarray}

\begin{figure}[btp]
\centering
\includegraphics[width=0.8\columnwidth]{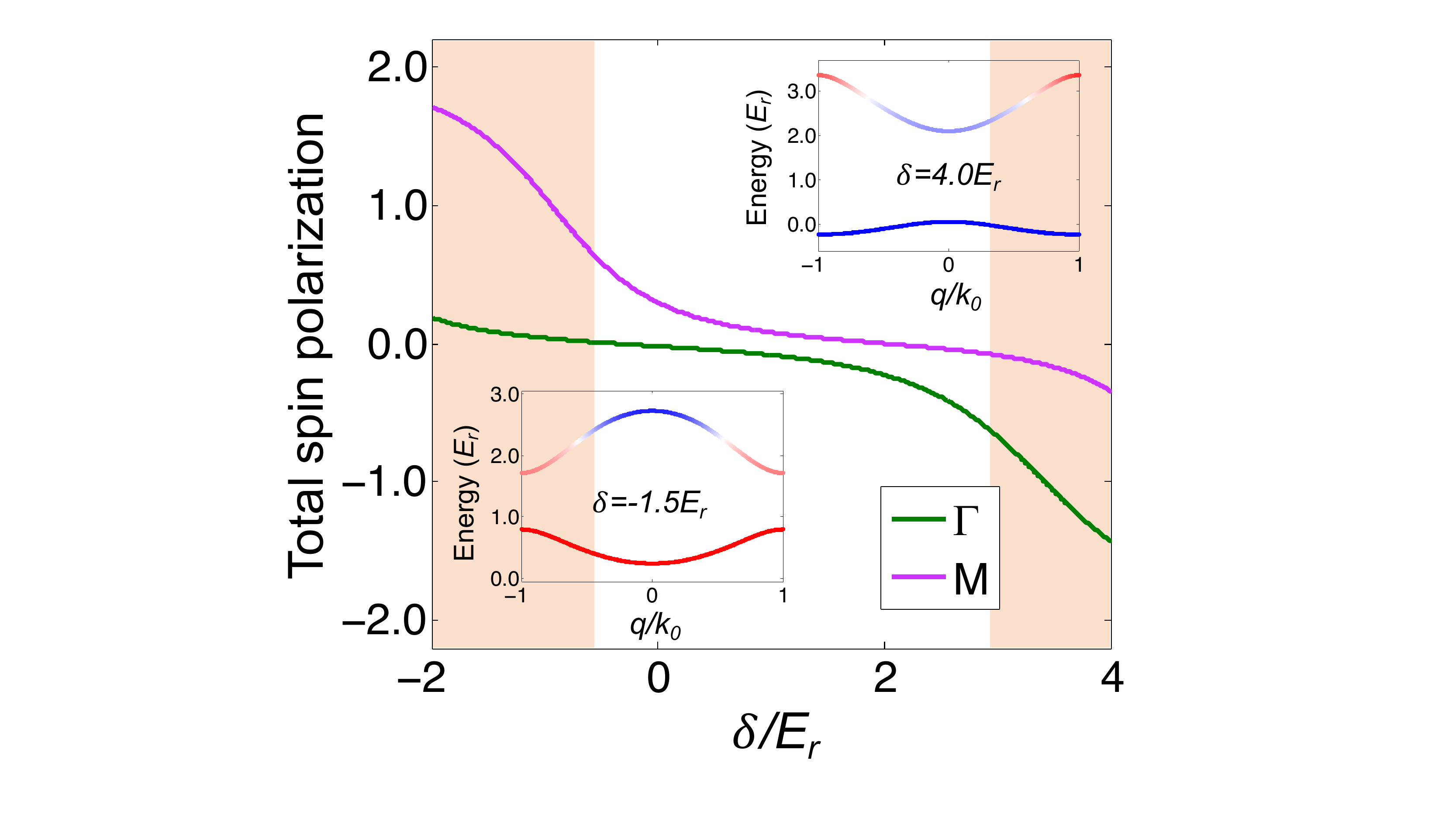}
\caption*{{\bf Figure S1 } Total spin polarizations of the lowest two bands at $\Gamma$ and $M$ points. Shaded regions are
where the hybridized of the upper band to $p$-orbits is relevant.
The intermediate region (white area) is in the regime where two-band model is quite valid.
Insets: Two examples of energy spectra in the deep shaded regimes.
Here $V_{\uparrow}=2.5E_{\rm r}$, $V_{\downarrow}=5E_{\rm r}$ and $M_0=1.0E_{\rm r}$.} \label{figs_sp}
\end{figure}

When the lattice is spin-dependent, e.g. $V_{\uparrow}=2.5E_{\rm r}$ and $V_{\downarrow}=5E_{\rm r}$, energy spectra of the
lowest two subbands, mainly the upper one, may be hybridized with $p$-orbits, as illustrated in the insets of Fig.~S1. Nevertheless,
one can find that in a broad parameter region
($\sim-0.5E_{\rm r}<\delta<3.0E_{\rm r}$ in Fig.~S1) the effect of the $p$-orbital band is negligible, in which regime the two-band model well capture the low band physics. Beyond this regime, the hybridization to $p$-orbital bands can be relevant.

\section{The $s$-band model}

Focusing on the $s$-band physics, one can write down the tight-binding model for the
Raman lattice Hamiltonian as
\begin{equation}
H_{\rm TB}=\sum_{\langle i,j\rangle,\sigma}(t_\uparrow^{ij}c^{\dagger}_{i\uparrow}c_{j\uparrow}+t_\downarrow^{ij}c^{\dagger}_{i\downarrow}c_{j\downarrow})
+\sum_{\langle i,j\rangle}(t_{\rm so}^{ij}c^\dagger_{j\uparrow}c_{j+1\downarrow}+{\rm h.c.})+\sum_im_z(n_{i\uparrow}-n_{i\downarrow}),
\end{equation}
where $c_{j\sigma}$ ($c_{j\sigma}^\dagger$) are the annihilation (creation) operators of $s$-orbit for
spin $\sigma=\uparrow,\downarrow$ at lattice site $j$, the atom number operator
$n_{j\sigma}\equiv c_{j\sigma}^\dagger c_{j\sigma}$, and the effective Zeeman term
$m_z=\delta/2-\Delta m_z$ is tuned by two-photon detuning, with $\Delta m_z$ being an onsite energy difference induced by spin-dependent lattice potential, i.e.
.
The spin-conserved hopping couplings $t_\sigma^{ij}$, which are generically spin-dependent due to the spin-dependent lattice potentials, the spin-flip hopping coefficients $t_{\rm so}^{ij}$, and the one-site term $\Delta m_z$ are respectively obtained by
\begin{equation}
t_\sigma^{ij}=\int dx\phi_{s\sigma}^{(i)}(x)\left[\frac{\hbar^2k_x^2}{2m}+V_{\sigma}^{\rm latt}(x)\right]\phi_{s\sigma}^{(j)}(x),
\end{equation}
\begin{equation}
t_{\rm so}^{ij}=\int dx\phi_{s\uparrow}^{(i)}(x){\cal M}(x)\phi_{s\downarrow}^{(j)}(x),
\end{equation}
\begin{equation}
\Delta m_z=\int dx \phi^{(0)}_{s\uparrow}(x)\left[\frac{\hbar^2k_x^2}{2m}+V_{\uparrow}\cos^2(k_0x)\right]
\phi^{(0)}_{s\uparrow}(x)-\int dx \phi^{(0)}_{s\downarrow}(x)\left[\frac{\hbar^2k_x^2}{2m}+V_{\downarrow}
\cos^2(k_0x)\right]\phi^{(0)}_{s\downarrow}(x),
\end{equation}
where $\phi_{s\sigma}^{(j)}(x)$ being the Wannier functions for $s$-bands. Since $\phi_{s\sigma}^{(j)}(x)=\phi_{s\sigma}^{(0)}(x-x_j)$ with $x_j=ja$ ($a$ is the lattice constant) and the anti-symmetry of the Raman potential ${\cal M}(x)$ with respect to each lattice site, we have
$t_\sigma^{ij}=-t_\sigma$ and $t_{\rm so}^{j,j\pm1}=\pm(-1)^jt_{\rm so}$, with
\begin{equation}
t_{\sigma}=-\int dx \phi^{(0)}_{s\sigma}(x)\left[\frac{\hbar^2k_x^2}{2m}+V_{\sigma}\cos^2(k_0x)\right]\phi^{(0)}_{s\sigma}(x-a),
\end{equation}
\begin{equation}
t_{\rm so}=M_0\int dx \phi^{(0)}_{s\uparrow}(x)\cos(k_0x)\phi^{(0)}_{s\downarrow}(x-a).
\end{equation}
The staggered sign in the spin-flip hopping coefficients $t_{\rm so}^{j,j\pm1}$ can be absorbed by a gauge transformation that $c_{j\downarrow}\to-e^{i\pi x_j/a}c_{j\downarrow}$~\cite{Liu2013b}. Further transforming
the tight-binding model into the Bloch momentum space yields
\begin{equation}
H_{\rm TB}=\sum_{q_x}\Psi_{q_x}^\dagger{\cal H}(q_x)\Psi_{q_x},
\end{equation}
where the spinor $\Psi_{q_x}\equiv(c_{q_x\uparrow},c_{q_x\downarrow})^{\sf T}$, with $c_{j\sigma}=\frac{1}{\sqrt{N}}\sum_{q_x} e^{iq_xx_j}c_{q_x\sigma}$ and
$N$ being the number of lattice sites, and the Bloch Hamiltonian
\begin{equation}
{\cal H}(q_x)=-2t_1\cos(q_xa)\otimes{\bf 1}+[m_z-2t_0\cos(q_xa)]\sigma_z-2t_{\rm so}\sin(q_xa)\sigma_y.
\end{equation}
Here we define $t_0\equiv(t_{\uparrow}+t_{\downarrow})/2$ and $t_1\equiv(t_{\uparrow}-t_{\downarrow})/2$.
Note that for a most generic investigation, we can take into account a Zeeman term in the form ${\vec m}=(m_x,m_y,m_z)$, which
represents the perturbations that may affect the symmetry of the Hamiltonian. We then have
\begin{equation}\label{Hq}
{\cal H}(q_x)=-2t_0\cos(q_xa)\sigma_z-2t_{\rm so}\sin(q_xa)\sigma_y-2t_1\cos(q_xa)\otimes{\bf 1}+{\vec m}\cdot{\vec\sigma}.
\end{equation}
Apart from the $m_z$ term, which is controlled by the two-photon detuning, the $m_x$ and $m_y$ terms can be, respectively, generated by Raman couplings
\begin{equation}\label{Mxy}
M_{0x}\sin(k_0x)\sigma_x,\quad{\rm and}\quad M_{0y}\sin(k_0x)\sigma_y,
\end{equation}
which induce only on-site spin flipping due to the symmetry of $\sin(k_0x)$ with respect to each lattice site. Thus we have
\begin{equation}
m_x=M_{0x}\int dx \phi^{(0)}_{s\uparrow}(x)\sin(k_0x)\phi^{(0)}_{s\downarrow}(x), \ \
m_y=M_{0y}\int dx \phi^{(0)}_{s\uparrow}(x)\sin(k_0x)\phi^{(0)}_{s\downarrow}(x).
\end{equation}

\section{Symmetries}

Before we solve the end states of the system, we analyze the symmetries of the Hamiltonian~\eqref{Hq}. We study in the following two different situations.

First, we consider the case that the lattice potential is spin-independent, $V_{\uparrow}^{\rm latt}=V_{\downarrow}^{\rm latt}$, in which regime the unity matrix term in Eq.~\eqref{Hq} vanishes $t_1=0$, and one can easily see that the bulk energy spectra of the Hamiltonian is symmetric. When the Zeeman coupling points in the $y-z$ plane, i.e. $m_x=0$, the Hamiltonian respects a chiral symmetry defined as the product of time-reversal symmetry ${\cal T}=i\sigma_y K$, with $K$ being complex conjugate, and the particle-hole symmetry which in the second quantization sends ${\cal C}: (c_{i,\sigma},  c_{i,\sigma}^\dag)\rightarrow(\sigma_z)_{\sigma\sigma'}(c_{i,\sigma'}^\dag,  c_{i,\sigma'})$. This renders the chiral transformation ${\cal S}: (c_{i,\sigma},  c_{i,\sigma}^\dag)\rightarrow(\sigma_xK)_{\sigma\sigma'}(c_{i,\sigma'}^\dag,  c_{i,\sigma'})$ and ${\cal S}H_{\rm TB}{\cal S}^{-1}=H_{\rm TB}$, while in the first quantization picture the chiral symmetry simply reads
\begin{equation}
{\cal S}=\sigma_x,
\end{equation}
and ${\cal S}{\cal H}(q_x){\cal S}^{-1}=-{\cal H}(q_x)$~\cite{Liu2013b}.
It is well known that this Hamiltonian belongs to the 1D AIII class in the AZ classification~\cite{AZ1997}, with the topology being characterized by the 1D winding number~\cite{Liu2013b,Ryu2010}.

More generically, as studied in the present work, we consider the spin-dependent lattice and then $t_{\uparrow}\neq t_{\downarrow}$, which gives a nonzero unity matrix term with $t_1\neq0$ in the Hamiltonian~\eqref{Hq}. In this case the usual locally defined chiral symmetry ${\cal S}$ is explicitly broken. Nevertheless, the novel hidden symmetries make the Hamiltonian be nontrivial. For convenience, we consider first the case with $\vec m=0$. In this case ${\cal H}(\vec m=0)$ satisfies a magnetic group symmetry defined as the product of time-reversal symmetry ${\cal T}=i\sigma_y K$ and mirror symmetry ${\cal M}_x=\sigma_x R_x$, giving
\begin{equation}
M_x={\cal T}{\cal M}_x=\sigma_zK\otimes R_x,
\end{equation}
with $M_x{\cal H}(q_x)M_x^{-1}={\cal H}(-q_x)$. Here $R_x$ denotes the spatial reflection along $x$ axis. On the other hand, the Hamiltonian also satisfies a {\it nonlocal chiral symmetry} defined as
\begin{equation}
{\cal S}=\sigma_z\otimes T_x(k_0)\otimes R_x,
\end{equation}
where $T_x(k_0)$ is a $k_0$-momentum translation, giving that ${\cal S}{\cal H}(q_x){\cal S}^{-1}=-{\cal H}(q_x)$. It is important that the above nonlocal chiral symmetry is fundamentally different from the usual locally defined counterpart, and it merely cannot protect nontrivial topology of the system. In the present generic case, the magnetic group symmetry is also necessary to reach the nontrivial topology phase. The midgap end states at left ($|\psi_{L}\rangle$) and right ($|\psi_R\rangle$) hand boundaries are transformed in the following way $M_x|\psi_{L/R}\rangle=|\psi_{R/L}\rangle$ and ${\cal S}|\psi_{L/R}\rangle=|\psi_{R/L}\rangle$. This is because the both symmetries include the spatial reflection operator $R_x$.
Moreover, the commutation and anti-commutation relations of the two symmetries with ${\cal H}$ imply that the magnetic group (nonlocal chiral) symmetry connects two states with identical (opposite) energy, given that the two states are independent. For this $|\psi_{L/R}\rangle$ have to be zero energy, i.e. midgap states in the presence of both symmetries. This is the case for $\vec m$ along $y$ axis. On the other hand, the $m_z\sigma_z$ term keeps $M_x$ symmetry, hence protects the end state degeneracy, while $m_x\sigma_x$ term breaks $M_x$ symmetry and splits out the degeneracy.

\section{Mid-gap edge states}

We consider the open boundary condition for the 1D lattice system, which are located at $x=0$ and $x=L$, and show first the result for $\vec m=0$.
The tight-binding Hamiltonian can be rewritten in the pseudospin-1/2 basis as
\begin{equation}
H_{\rm TB}=\sum_j({\cal A}c_j^\dagger c_{j+1}+{\cal A}^\dagger c_{j+1}^\dagger c_{j}),
\end{equation}
where
\begin{equation}\label{def_A}
{\cal A}=-t_0\sigma_z+it_{\rm so}\sigma_y-t_1{\bf 1}.
\end{equation}
The gap states induced by hard-wall boundary are exponentially localized around $x=0,L$ with the wavefunctions taking the form of ansatz
\begin{equation}\label{ansatz}
|\psi_{\rm L/R}(x_j)\rangle=\frac{1}{\sqrt{{\cal L}}}\lambda^{x_j/a}|\varphi_{\rm L/R}\rangle,
\end{equation}
where $\lambda$ is a complex number, ${\cal L}$ is the normalization factor, and $|\varphi_{\rm L/R}\rangle$ are two-component spinors.
Since the above wavefunctions decay (or increase) with $x_j$ when $|\lambda|<1$ (or $|\lambda|>1$), the gap state localized on the left (right) boundary
corresponds to a solution with $|\lambda|<1$ ($|\lambda|>1$). From $H_{\rm TB}|\psi\rangle=E|\psi\rangle$, it follows that
\begin{equation}
(\lambda{\cal A}+\lambda^{-1}{\cal A}^\dagger)|\varphi_{\rm L/R}\rangle=E|\varphi_{\rm L/R}\rangle,
\end{equation}
which gives the energies
\begin{equation}
E_{\pm}(\lambda)=-t_1(\lambda+\lambda^{-1})\pm\sqrt{t_0^2(\lambda+\lambda^{-1})^2-t_{\rm so}^2(\lambda-\lambda^{-1})^2}.
\end{equation}
Since $E_{\pm}$ should lie inside the band gap, i.e. $E_{\pm}(\lambda)<\Delta_c$ holds for any infinitesimal $t_{\rm so}$,
the zero energy solutions are supposed to be valid. By setting $E_{\pm}(\lambda)=0$, we have the solutions
\begin{equation}
\lambda_{\rm L/R}^{\pm}=\mp i\lambda_{\rm L/R},
\end{equation}
where
\begin{equation}
\lambda_{\rm L}=\left(\frac{\sqrt{t_0^2-t_1^2}-t_{\rm so}}{\sqrt{t_0^2-t_1^2}+t_{\rm so}}\right)^{1/2},\quad \lambda_{\rm R}=1/\lambda_{\rm L}.
\end{equation}
Since $\lambda_{\rm L}<1$, the corresponding eigen-state represents the spinor of edge state on the left boundary
\begin{equation}\label{leftend}
|\varphi_{\rm L}\rangle=\frac{1}{\sqrt{1+t_2}}\binom{1}{-\sqrt{t_2}},
\end{equation}
where $t_2\equiv t_\uparrow/t_{\downarrow}$. The other solution $\lambda_{\rm R}>1$ gives
\begin{equation}\label{rightend}
|\varphi_{\rm R}\rangle=\frac{1}{\sqrt{1+t_2}}\binom{1}{\sqrt{t_2}}.
\end{equation}
Due to the hard-wall boundary condition, i.e. $|\psi_{\rm L}(x_j=0)\rangle=0$ and $|\psi_{\rm R}(x_j=L)\rangle=0$, the zero-energy (mid-gap) edge states
should be superposition of the $\lambda_{\rm L/R}^{\pm}$ solutions
\begin{equation}\label{edgeL0}
|\psi_{\rm L}(x_j)\rangle=\frac{1}{\sqrt{\cal L}}\lambda_{\rm L}^{x_j/a}\left(e^{i\pi x_j/2a}-e^{-i\pi x_j/2a}\right)|\varphi_{\rm L}\rangle
\end{equation}
and
\begin{equation}\label{edgeR0}
|\psi_{\rm R}(x_j)\rangle=\frac{1}{\sqrt{\cal L}}\lambda_{\rm R}^{(x_j-L)/a}\left[e^{i\pi (x_j-L)/2a}-e^{-i\pi (x_j-L)/2a}\right]|\varphi_{\rm R}\rangle.
\end{equation}

We further consider the effect of $m_y\sigma_y$ term on the edge states. Supposing the ansatz in Eq.~(\ref{ansatz}) is still applicable, we
have the following eigenvalue equation
\begin{equation}
\left[h(\lambda)-E\right]|\varphi\rangle=0,
\end{equation}
where
\begin{equation}
h(\lambda)\equiv\lambda{\cal A}+\lambda^{-1}{\cal A}^\dagger+m_y\sigma_y,
\end{equation}
with ${\cal A}$ given in Eq.~(\ref{def_A}).
Due to the hard-wall boundary condition, there must be two different numbers $\lambda_{1,2}$ giving the same energy $E(\lambda_{1})=E(\lambda_{2})$
and also the same spinor state $\left[h(\lambda_{1,2})-E\right]|\varphi\rangle=0$ for each edge state, so that edge states can take
similar forms to Eqs.~(\ref{edgeL0}-\ref{edgeR0}). The eigenvalue equation gives ($\alpha=1,2$)
\begin{equation}
\left[ \begin{array}{cc}
-(t_0+t_1)(\lambda_\alpha+\lambda_\alpha^{-1})-E & t_{\rm so}(\lambda_\alpha-\lambda_\alpha^{-1})-im_y  \\
-t_{\rm so}(\lambda_\alpha-\lambda_\alpha^{-1})+im_y  &  (t_0-t_1)(\lambda_\alpha+\lambda_\alpha^{-1})-E
\end{array} \right]\binom{\varphi_\uparrow}{\varphi_\downarrow}=0.
\end{equation}
Thus we have the following two equations ($\varphi_\uparrow/\varphi_\downarrow$)
\begin{equation}
\left\{\begin{array}{l}
\frac{-(t_0+t_1)(\lambda_1+\lambda_1^{-1})-E}{t_{\rm so}(\lambda_1-\lambda_1^{-1})-im_y}=\frac{-(t_0+t_1)(\lambda_2+\lambda_2^{-1})-E}{t_{\rm so}(\lambda_2-\lambda_2^{-1})-im_y},
\\
\frac{-t_{\rm so}(\lambda_1-\lambda_1^{-1})+im_y}{(t_0-t_1)(\lambda_1+\lambda_1^{-1})-E}=\frac{-t_{\rm so}(\lambda_2-\lambda_2^{-1})+im_y}{(t_0-t_1)(\lambda_2+\lambda_2^{-1})-E},
\end{array}\right.
\end{equation}
which lead to
\begin{equation}
\frac{(t_0+t_1)(\lambda_1+\lambda_1^{-1})+E}{(t_0+t_1)(\lambda_2+\lambda_2^{-1})+E}=\frac{(t_0-t_1)(\lambda_1+\lambda_1^{-1})-E}{(t_0-t_1)(\lambda_2+\lambda_2^{-1})-E}.
\end{equation}
It is obvious that $E=0$ is the valid solution. Therefore, we have demonstrated that perturbation like $m_y\sigma_y$ do not change the zero energy of the two edge states and preserve both the mirror symmetry and nonlocal chiral symmetry. On the other hand, from the solutions~\eqref{leftend} and~\eqref{rightend} one can see that the end states are both polarized to $+z$ (or $-z$) direction if $|t_2|<1$ (or $|t_2|>1$), while are oppositely polarized along $x$ direction. Thus the Zeeman perturbation term $m_z\sigma_z$ still keeps the degeneracy of the end states, while $m_x\sigma_x$ splits out the degeneracy, consistent with the symmetry analysis given in the previous section.

\begin{figure}[btp]
\centering
\includegraphics[width=0.45\columnwidth]{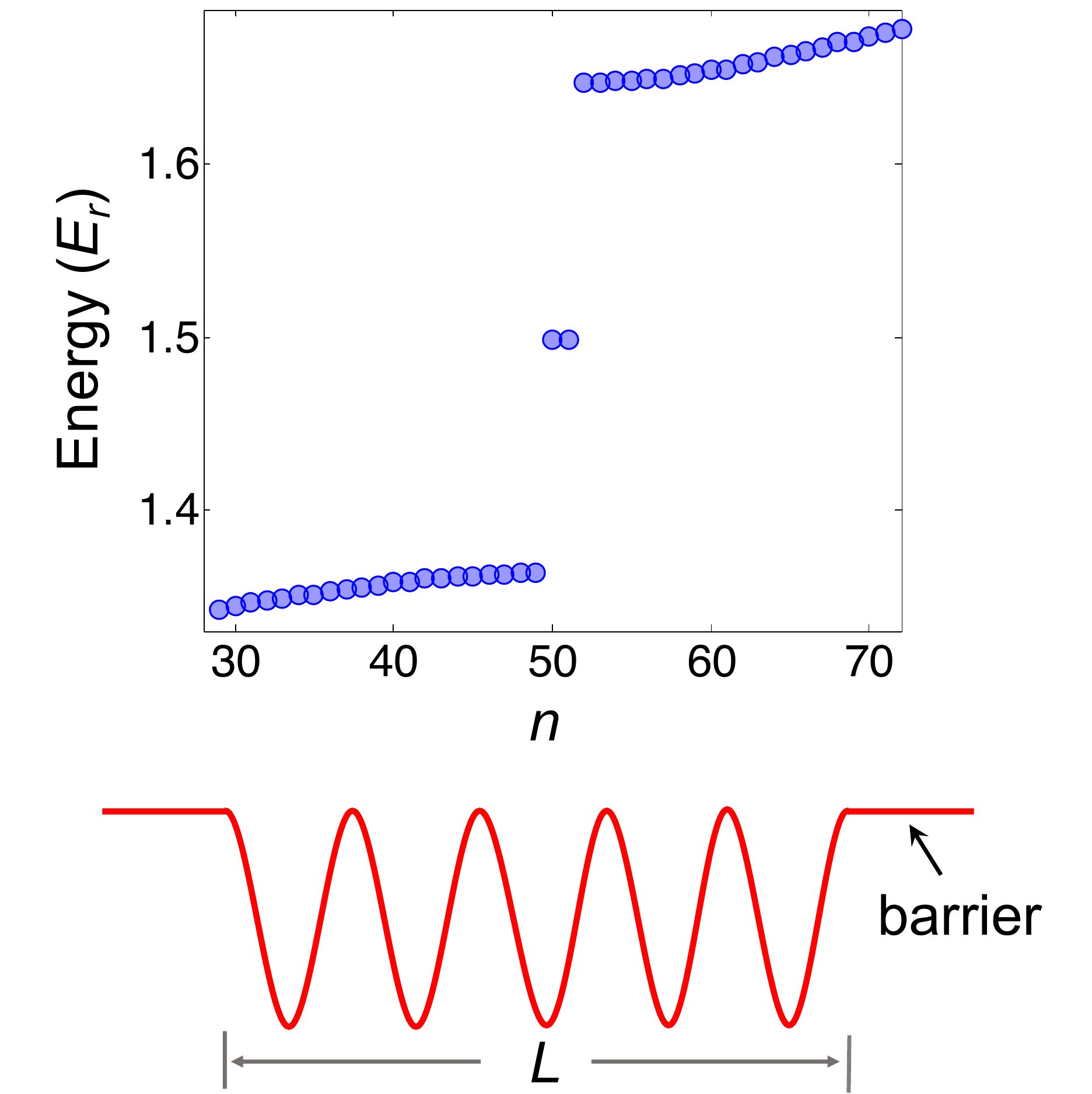}
\caption*{{\bf Figure S2 } Energy spectra of 1D Raman lattice Hamiltonian $H$, given by Eq.~(1) of main text, with open boundary condition, showing
the existence of mid-gap end states. The results are obtained by full band diagonalization without $s$-band approximation.
The parameters take $V_{\uparrow}=2.5E_{\rm r}$, $V_{\uparrow}=5E_{\rm r}$, $M_0=1E_{\rm r}$,
and $\delta=0.679E_{\rm r}$ (the effective Zeeman shift $\vec m=0$).
} \label{figs_edge}
\end{figure}

When the $s$-band approximation is valid, the original 1D Raman lattice Hamiltonian $H$ [Eq.~(1) in the main text] without Zeeman coupling tem gives two mid-gap end states under the open boundary condition (Fig.~S2). Exact full band diagonalization in the real space are used to calculate all the eigenvalues. The open boundary condition is constructed in the way that at the end of lattice two finite barriers are added with the same height as the lattice (see Fig.~S2). One can also construct the boundary by adding a large Zeeman term along $y$ axis at the end of the lattice. We can further examine the effect of the Zeeman perturbation field ${\vec m}\cdot{\vec \sigma}$
by tuning the two-photon detuning $\delta$ to induce $m_z$ or adding on-site Raman couplings to generate $m_{x,y}$, as given in Eq.~(\ref{Mxy}).

\section{Quench dynamics}

To consider the dynamics of generally mixed states after a sudden quench,
we deal with the time-dependent density matrix $\rho(t)$ and the
master equation of the so-called Lindblad form [Eq.~(2) in the main text].
The master equation takes into account the dissipations that
exist in cold atom experiments and can be caused by laser noise or unstable magnetic field.

\begin{figure}[btp]
\centering
\includegraphics[width=0.6\columnwidth]{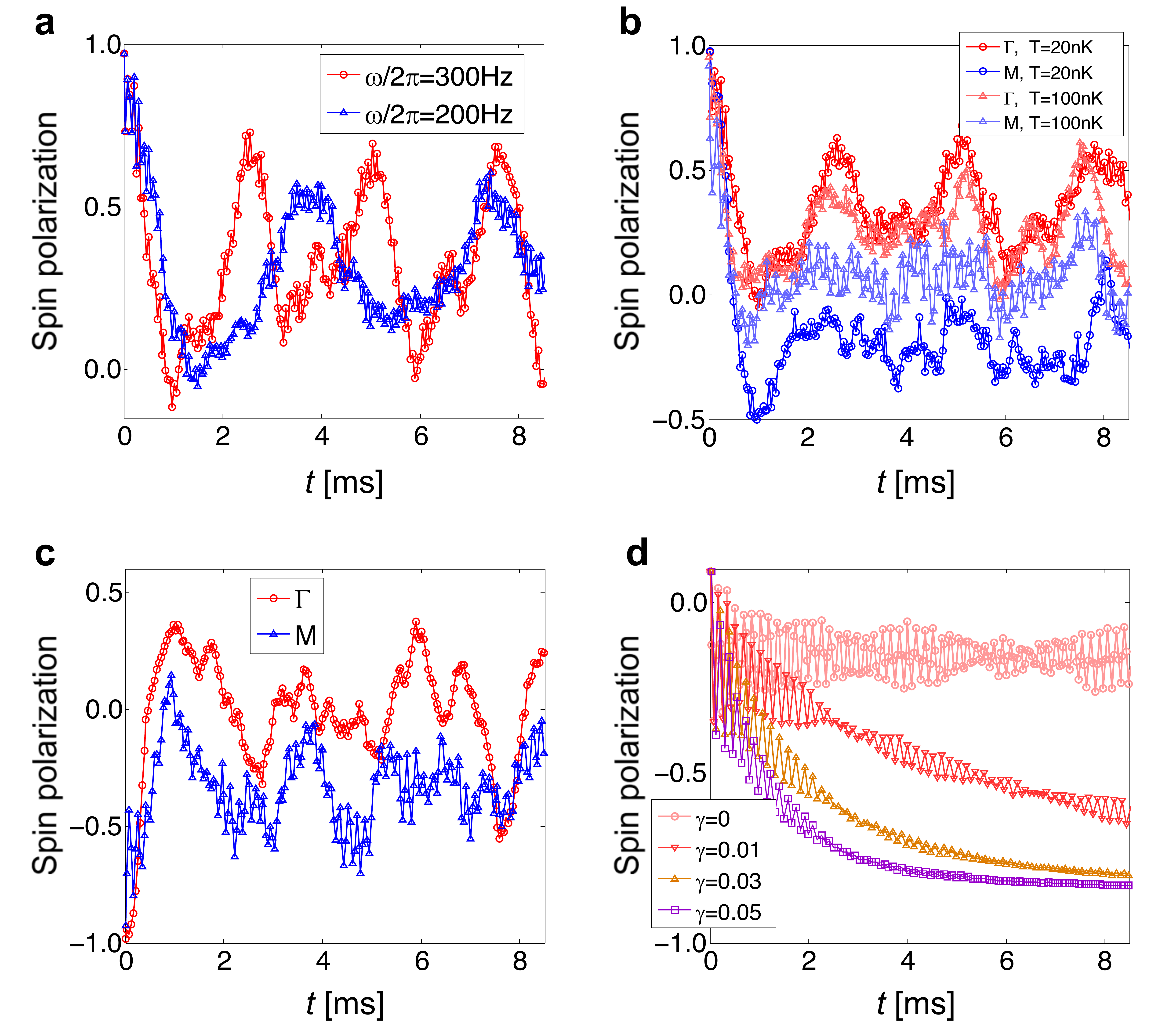}
\caption*{{\bf Figure S3 } Quench dynamics under various conditions. {\bf a,} The time-evolution of spin polarizations at the $\Gamma$  point
with different trapping frequencies after a quench from $\delta_i=-2E_{\rm r}$ (trivial)
to $\delta_f=1E_{\rm r}$ (topological). The trapping frequency governs the period of collective oscillation.
{\bf b,} The spin dynamics at different temperature with $\delta_i=-2E_{\rm r}$
and $\delta_f=1E_{\rm r}$. Temperature tends to smooth down the oscillation and spin polarization.
{\bf c,} The quench dynamics from $\delta_i=2.5E_{\rm r}$ (trivial) to $\delta_f=1E_{\rm r}$ (topological) shows a
clearly different behavior compared to the converse quench process from $\delta_i=1E_{\rm r}$ to $\delta_f=2.5E_{\rm r}$ (see {\bf d} or Fig.~2b in the main text).
{\bf d,} The spin dynamics at the $\Gamma$ point after a quench from $\delta_i=1E_{\rm r}$ to $\delta_f=2.5E_{\rm r}$ with different decay rates at $T=130$nK.
In (b-d), the trapping frequency is set at $\omega_x=(2\pi)300$Hz. The quantum spin dynamics show sharp difference in quenching to topological and trivial phases. In all the cases, we have considered here $V_{\uparrow}=1E_{\rm r}$, $V_{\downarrow}=4E_{\rm r}$ and $M_0=1E_{\rm r}$.
} \label{figs_quench}
\end{figure}

Regardless of the dissipation, the time evolution is unitary
\begin{equation}
\rho(t)=U(t)\rho(0)U(t)^\dagger,
\end{equation}
where the evolution operator $U(t)=e^{-i[H(\delta_f)+V_{\rm trap}]t}$ with the harmonic trap
$V_{\rm trap}=\frac{1}{2}m\omega_xx^2$.
There are two mechanisms governing the quench
dynamics: (1) Nonequilibrium population after a quench causes interband oscillations at individual momenta in the first Briliouin zone (FBZ), which can be regarded as
the precessional motion for each momentum-linked spin. (2) The external trapping potential induces intraband coupling between Bloch states at different momenta.
To be specific, we first consider the sudden quench from trivial to
topological regime. To capture the key physics in the theory, we assume that the dynamics are dominated by quantum states of the lowest two subbands before and
after the quench (note that the numerical simulation is however based on full band study). One can then approximate the effective Hamiltonian for quench dynamics as
\begin{equation}
H_{\rm eff}\sim\sum_{q}E_{n_0}(q_x)|\Psi_{n_0}(q_x)\rangle\langle \Psi_{n_0}(q_x)|\propto\sum_{q}\left[B_x(q_x)\sigma_x+B_z(q_x)\sigma_z\right],
\end{equation}
where the quantum number $n_0$ labels the lowest band and $B_{x,z}(q_x)$ are the components of the effective momentum-dependent magnetic field.
The spin texture (e.g. the third row in Fig.~1c of the main text) implies
that the post-quench Hamiltonian $H(\delta_f)$ assumes a rapidly varying magnetic field if $\delta_f$ enters the topological regime.
In the trivial regime one has that $B_z(q_x)>0$ (or $B_z(q_x)<0$) for the entire FBZ, rendering an averagely $z$ directional magnetic field. In contrast, the effective magnetic field $\vec B$ winds over all direction in $x-z$ plane in the topological phase. In particular, at the $\Gamma$ and $M$ points $B_z(0)$ and $B_z(\pi)$ are opposite. This feature has crucial effect on the quench dynamics induced by the trapping driven intraband transitions.
Specifically,
the trapping potential correlates the spins and induces spin-wave-like collective oscillation,
with the oscillation period being determined by the trapping frequency $\omega_x$ (Fig.~S3a).
Note that the oscillation frequency in Fig.~S3a is a bit larger than $\omega_x$, which can be explained as the consequence
of off-resonance in the intraband coupling induced by trapping potential. The oscillation frequency can be estimated as $\sqrt{\omega_x^2+\Delta E^2}$, where
$\Delta E$ denote the energy difference between the $\Gamma$ and $M$ points.
On the other hand, the quench from topologically nontrivial to trivial regime, however, results in different dynamical behaviors,
since the post-quench magnetic field is nearly uniform in the FBZ and collective dynamics is negligible.
The crucial role of the post-quench band topology (effective magnetic field) can be reflected by the big difference between the dynamical behavior caused by
a quench from $\delta_i=1E_{\rm r}$ to $\delta_f=2.5E_{\rm r}$ (Fig.~2b in the main text) and the one due to a converse quench process (Fig.~S3c).

In our calculations, the initial state is given by $\rho(0)=\sum_np_n|n\rangle\langle n|$.
Here, $|n\rangle$ are the eigenstates of $H(\delta_i)+V_{\rm trap}$, with $p_n$ being the corresponding
occupation probability, which is determined by Fermi-Dirac statistics.
At zero temperature, we set $p_n=\Theta(\mu-E_n)$. Here, $\Theta(\cdot)$ denotes the unit step function
and the chemical potential (Fermi energy) $\mu$ is assumed just above the
maximum energy of the lowest band. Note that the ``lowest band" of the Hamiltonian $H(\delta_i)+V_{\rm trap}$ which has trapping potential actually refers to the corresponding energy band excluding the trap, for which
a proper chemical potential $\mu$ can be chosen. For instance, in the case with $\delta_i=-2E_{\rm r}$ and $\delta_f=1E_{\rm r}$
[Fig.~2(a,c) in the main text and Fig.~S3(a-b)], we set the chemical potential $\mu=0.3E_{\rm r}$.
At finite temperature, for comparison, we set the same chemical potential $\mu$ as at $T=0$.
It can be shown that temperature tends to smooth down the oscillations (Fig.~S3b), but
at low enough $T$, the dynamical behaviors still have similar characteristics to the zero-temperature case .

The noise-induced dissipation (or dephasing) can be qualitatively captured by the non-unitary part of the Lindblad master equation,
where the operator $L$ takes the form $L=(\widetilde{\sigma}_x+i\widetilde{\sigma}_y)/2$ (or $L=\widetilde{\sigma}_z$),
describing the decay of excitations (or phase coherence) at a rate $\gamma$. Here $\widetilde{\sigma}_{x,y,z}$ denote the
Pauli matrices in the eigenbasis of the post-quench $H(\delta_f)+V_{\rm trap}$:
\begin{equation}
\widetilde{\sigma}_{x,y,z}=\sum_{n_1,n_2}\langle n_1|\Sigma_{x,y,z}|n_2\rangle|n_1\rangle\langle n_2|,
\end{equation}
with
\begin{equation}
\Sigma_{j}\equiv\underbrace{\sigma_{j}\otimes\sigma_{j}\otimes\cdots\otimes\sigma_{j}}_{N}=\bigotimes_{N}\sigma_{j}\quad(j=x,y,z).
\end{equation}
In our calculations,
dephasing and decay are, respectively, considered
in the quench dynamics from trivial to topological regime and the other way round.  In the latter case,
the spin polarization can approach gradually the post-quench spin texture by relaxation at different rates that depend on the environment (Fig.~S3d).

\begin{figure}
\centering
\includegraphics[width =0.6\textwidth]{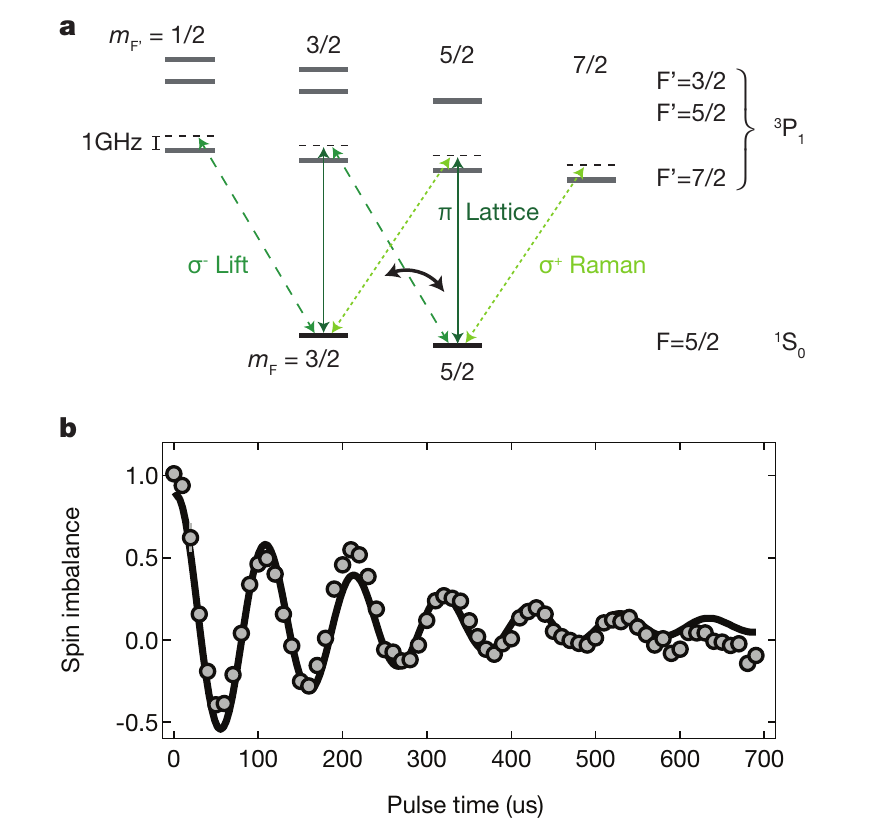}
\caption*{{\bf Figure S4 } Electronic hyperfine level of $^{173}$Yb atoms and Raman coupling. {\bf a},The degeneracy of $^1S_0$ ground-state is lifted by a $\sigma^-$ polarized light (called lift beam) propagating along $\hat{x}$-direction. An optical AC Stark effect then separates out an effective spin-$\frac{1}{2}$ subspace from other hyperfine levels for the realization of effective spin-$\frac{1}{2}$ SOC. The optical lattice beam has polarization of $\pi$, which acts as one laser beam to establish the the Raman coupling together with a third laser with polarization of $\sigma^+$. All the three laser have the same frequency $\sim$1$\text{GHz}$ blue detuned from the transition between $|F=5/2\rangle$ and $|F'=7/2\rangle$. {\bf b}, Spin dynamics during the Rabi oscillation. An incident beam of the lattice together with the Raman beam couples the $m_F=$5/2 and $m_F=$3/2 states inducing the oscillation of spin imbalance $(N_{\uparrow}-N_{\downarrow})/(N_{\uparrow}+N_{\downarrow})$ where $N_{\uparrow} (N_{\downarrow})$ is the number of spin-up (spin-down) atoms.  The effective Rabi frequency $\Omega_{eff}$ is calibrated by fitting the damped spin oscillation. }
\label{fig:supp1energylevels}
\end{figure}

\section{1D Optical lattice and periodic Raman coupling potential}
 A spin-dependent optical lattice potential is created by an incident light and its reflected light from a mirror along $ x $ direction, $ \textbf{E}_{1x} = \textbf{e}_{z} 2\overline{E}_{1x} e^{i(\varphi_{1x}+\varphi'_{1x})/2}\text{cos}(k_{0}x +\phi_0) $, where $ \overline{E}_{1x} $ is the amplitude of the field, $ \varphi_{1x} $ ($ \varphi'_{1x} $) is the optical phase of the incident (retro-reflected) light at the position of atoms,  and $\phi_0=(\varphi_{1x}-\varphi'_{1x})/2$. The electric field of the Raman and the lift beam are plane-wave fields along $ z $ direction, $ \textbf{E}_{1z} = (\textbf{e}_{x} + i\textbf{e}_{y})\overline{E}_{1z} e^{i(k_{0}z+\varphi_{1z})} /\sqrt{2} $ and $ \textbf{E}_{2z} = (\textbf{e}_{x} - i\textbf{e}_{y})\overline{E}_{2z} e^{i(k_{0}z+\varphi_{2z})} /\sqrt{2} $, respectively.

In the diagram of energy levels shown in Fig.~S4, due to the single photon detuning of the light in the same order of the hyperfine splitting of related $^{173}\text{Yb}$ excited states, the potentials are taken into account all the three transitions from $ \lvert F=5/2\rangle $ to $ \lvert F'=3/2\rangle $, $ \lvert F'=5/2\rangle $, and $ \lvert F'=7/2\rangle $ in $ 6s^{2} $ $ ^{1}$S$_{0} $ $ \rightarrow $ $6s6p$ $ ^{3}$P$_{1} $. Therefore the lattice trap depth for $ \lvert \uparrow\rangle\equiv\lvert F=5/2, m_{F}=5/2\rangle $ and $ \lvert \downarrow \rangle\equiv\lvert F=5/2, m_{F}=3/2\rangle $ are given as,
\begin{equation}
V_{\sigma} = \sum_{F'} \frac{\hbar}{4\Delta_{F'}} |\Omega_{\sigma,F'}^{1x}|^{2} = \frac{\hbar}{4\Delta_{3/2}} |\Omega_{\sigma, 3/2}^{1x}|^{2} + \frac{\hbar}{4\Delta_{5/2}} |\Omega_{\sigma, 5/2}^{1x}|^{2} + \frac{\hbar}{4\Delta_{7/2}} |\Omega_{\sigma,7/2}^{1x}|^{2}
\end{equation}
where  $ \Delta_{F'} $ and $ \Omega_{\sigma, F'}^{1x} $ are the single photon detuning  and the Rabi frequency respectively for the transition from  $ \lvert \sigma \rangle $ to $ \lvert F'\rangle $ with $\sigma=\{\uparrow,\downarrow\}$. The Rabi frequency $ \Omega_{\sigma,F'}^{1x} $ is determined by $ \langle \sigma \lvert ez \lvert F',m_{F'}\rangle \textbf{e}_{z} \cdot \textbf{E}_{1x} /\hbar$ and only $ m_{F'} = m_{F} $ is considered here according to the selection rule. Similarly, the later Rabi frequency for $ \sigma^{+} $ transition Raman light and $ \sigma^{-} $ transition lift light are only considered correspondingly $ m_{F'} = m_{F} +1 $ and $ m_{F'} = m_{F} - 1 $. The spin-dependent lattice potential therefore reads,
$V_{\sigma}^{\rm latt}(x) = V_{\sigma}\cos^2 (k_{0}x+\phi_0)$ where the spin-dependent lattice depth is
\begin{equation}
V_{\sigma}= \sum_{F'} \frac{\lvert \langle \sigma \lvert ez \lvert F',m_{F}\rangle\lvert^{2}}{\hbar \Delta_{F'}} \overline{E}_{1x}^{2}
\end{equation}
Similarly, the energy shift by a lift beam $ U_{\sigma}^{\rm lift} $ is,
\begin{equation}
U_{\sigma}^{\rm lift} = \sum_{F'} \frac{\hbar}{4\Delta_{F'}} |\Omega_{\sigma, F'}^{2z}|^{2} = \sum_{F'} \frac{\lvert \langle \sigma \lvert e(x+y)/\sqrt{2} \lvert F',m_{F}-1\rangle\lvert^{2}}{4\hbar \Delta_{F'}} \overline{E}_{2z}^{2}
\end{equation}

Here the frequency difference between lift beam and Raman lattice beam is omitted, as it is much smaller than $ \Delta_{F'} $.
The Raman coupling potential are driven by the light $ \textbf{E}_{x} $ and $ \textbf{E}_{1z} $. Considering all the transitions, we obtain the Raman potential,
\begin{equation}
\mathcal{M}(x) = M_{0}\text{cos}(k_{0}x+\phi_0)e^{i(k_{0}z+\varphi_{1z} -\varphi_{1x}/2-\varphi'_{1x}/2)}
\end{equation}
where the amplitude of Raman potential becomes,
\begin{equation}
M_{0} = \sum_{F'} \frac{\hbar \Omega_{\uparrow,F'}^{1x^{*}} \cdot \Omega_{\downarrow,F'}^{1z} }{2\Delta_{F'}} = \sum_{F'} \frac{ \langle \sigma \lvert ez \lvert F',m_{F}\rangle \langle \sigma \lvert e(x+y)/\sqrt{2} \lvert F',m_{F}+1\rangle}{\hbar \Delta_{F'}} \overline{E}_{1x}\overline{E}_{1z}
\end{equation}
Since the quantum dynamics along $x$ direction is decoupled to other directions, we  set $z=0$ for the irrelevant phase $e^{ik_0 z}$ and gauge out the phase $e^{i(\varphi_{1z} -\varphi_{1x}/2-\varphi'_{1x}/2)}$. Furthermore we can set $\phi_0=0$ by redefining the origin of the coordinate.
Therefore, the final effective Hamiltonian reads,
\begin{equation}
H = [\frac{\hbar^{2}\textbf{k}^2}{2m}+\frac{V_{\uparrow}^{\rm latt}(x)+V_{\downarrow}^{\rm latt}(x)}{2}] \otimes \hat{\textbf{1}} + [m_{0z} + \frac{V_{\uparrow}^{\rm latt}(x)-V_{\downarrow}^{\rm latt}(x)}{2}]\sigma_{z} + \mathcal{M}(x)\sigma_{x}
\end{equation}
with $ m_{0z} = (\delta  + U_{\uparrow}^{\rm lift} - U_{\downarrow}^{\rm lift})/2$, $\mathcal{M}(x) = M_{0}\text{cos}(k_{0}x)$ and $V_{\sigma}^{\rm latt}(x) = V_{\sigma}\cos^2(k_{0}x)$ without loss of generality. Here $ \delta $ is the two-photon detuning of the Raman process.

\section{Three-level model}

In experiments, we exploit the $F=5/2$ manifold hyperfine levels of $^{173}$Yb and construct an effective two-level system consisting of
the states $|F=5/2,m_F=5/2\rangle$ and $|F=5/2,m_F=3/2\rangle$ (Fig.~\ref{fig:supp1energylevels}) by separating
other unwanted hyperfine levels away by AC Stark shifts. However, the existence of redundant states will more
or less modify the observations.  Hence, we consider a three-level system with one more level
$|F=5/2,m_F=1/2\rangle$ and expect the calculations to be more agreeable with experimental measurements.
By calculating the Clebsch-Gordan (CG) coefficients, we have the three-level Hamiltonian
\begin{equation}
H_{\rm 3l}=\frac{\hbar^2k_x^2}{2m}\otimes{\bf \hat 1}_3
+\left( \begin{array}{ccc}
V_{5/2}\cos^2k_0x+m_z & M_0\cos k_0x & 0 \\
M_0\cos k_0x & V_{3/2}\cos^2 k_0x-m_z & M'_0\cos k_0x \\
0 & M'_0\cos k_0x  & V_{1/2}\cos^2 k_0x-3m_z+\varepsilon
\end{array} \right),
\end{equation}
where ${\bf 1}_3$ is the 3-by-3 unit matrix, the lattice depth $V_{3/2}=3.84V_{5/2}$ and $V_{1/2}=5.28V_{5/2}$, Raman coupling
strength $M_0'=1.036M_0$, and $\varepsilon$ denotes the quadratic Zeeman shift, which is set to be $1.72E_{\rm r}$ here.

The pre-quench occupation for each eigenstate of $H_{\rm 3l}+V_{\rm trap}({\bf r})$, where $V_{\rm trap}({\bf r})=\sum_{i=x,y,z}\frac{1}{2}m\omega_i^2r_i^2$
(the trap frequencies are $\{\omega_x,\omega_y,\omega_z\}=2\pi\times\{260,190,50\}$Hz in our setup),
is given by the Fermi-Dirac distribution, with the chemical potential $\mu$ being determined by the total atom number $N_0\approx10^4$.
Based on the three-level Hamiltonian $H_{\rm 3l}$, we calculate the spin polarizations varying with the detuning $\delta$ at temperature $T\simeq 100$nK
(the inset of Fig.~4). The implied topological regime agrees with the experimental measurements in Fig.~4 of the main text.

\begin{figure}
	\centering
	\includegraphics[width =0.6\textwidth]{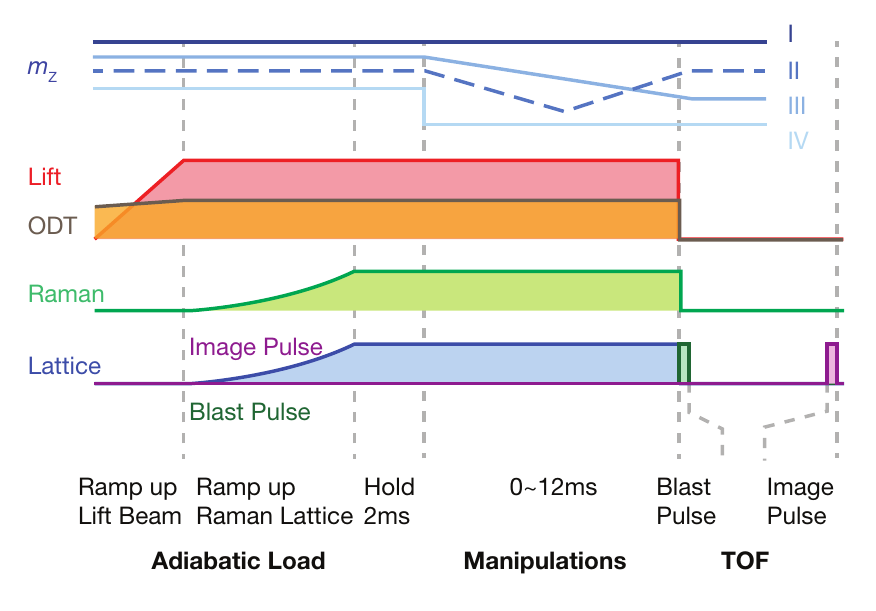}
	\caption*{{\bf Figure S5 } Experimental sequences. After evaporative cooling, we first inearly ramp up lift beam within 3~ms and adiabatically load atoms into the Raman lattice within 10~ms followed by 2 ms hold. In the stage of manipulation, we carry out  three measurements via controlling the two photon detuning $\delta$  (or $ m_{z}) $, including (I) the measurement of the $Z_2$ invariants, (II) adiabatic control of band topology, (IV) the far-from-equilibrium spin dynamics after quench between topologically distinct phases. The adiabaticity of the $\delta$-sweep as shown in the inset of Figure 4e is investigated with the sequence (III). A blast pulse is applied before the time-of-flight expansion, followed by the absorption imaging.}
	\label{fig:supp2sequence}
\end{figure}

\section{Calibration of the optical Raman lattice}
The effective Rabi coupling $\Omega_{\rm eff}$ between $ \lvert \uparrow\rangle $ and $ \lvert \downarrow\rangle $ states is calibrated by monitoring the spin dynamics when a brief pulse of the incident lattice light and the Raman beam is switched onto a spin-polarized gas. As shown in Fig.~S4,  the total spin imbalance, defined as $(N_{\uparrow}-N_{\downarrow})/(N_{\uparrow}+N_{\downarrow})$, exhibits a damped oscillation when a nonuniform spin-orbit gap is opened, which is the consequence of momentum-dependent spin-orbit coupling in the system. The strength of Raman coupling $M_0$ in the optical Raman lattice is then given by $M_0=2\Omega_{\rm eff}$.

Next the depth of the optical lattice is independently calibrated by means of the {\it in-situ} amplitude-modulation spectroscopy that probes the energy gap between the first and the third lowest-energy bands. To avoid spin-dependent lattice effect and the interaction-induced shift, a degenerate spin-polarized $|m_F=\frac{5}{2}\rangle$ gas is prepared in the lattice without the Raman and the lift beam. We note that the change of the lattice depth induced by the Raman and the lift beam is negligible. In addition the energy level shift induced by the AC Stark shift is spectroscopically calibrated.

\section{Experimental sequence}

\begin{figure}
	\centering
	\includegraphics[width =0.95\textwidth]{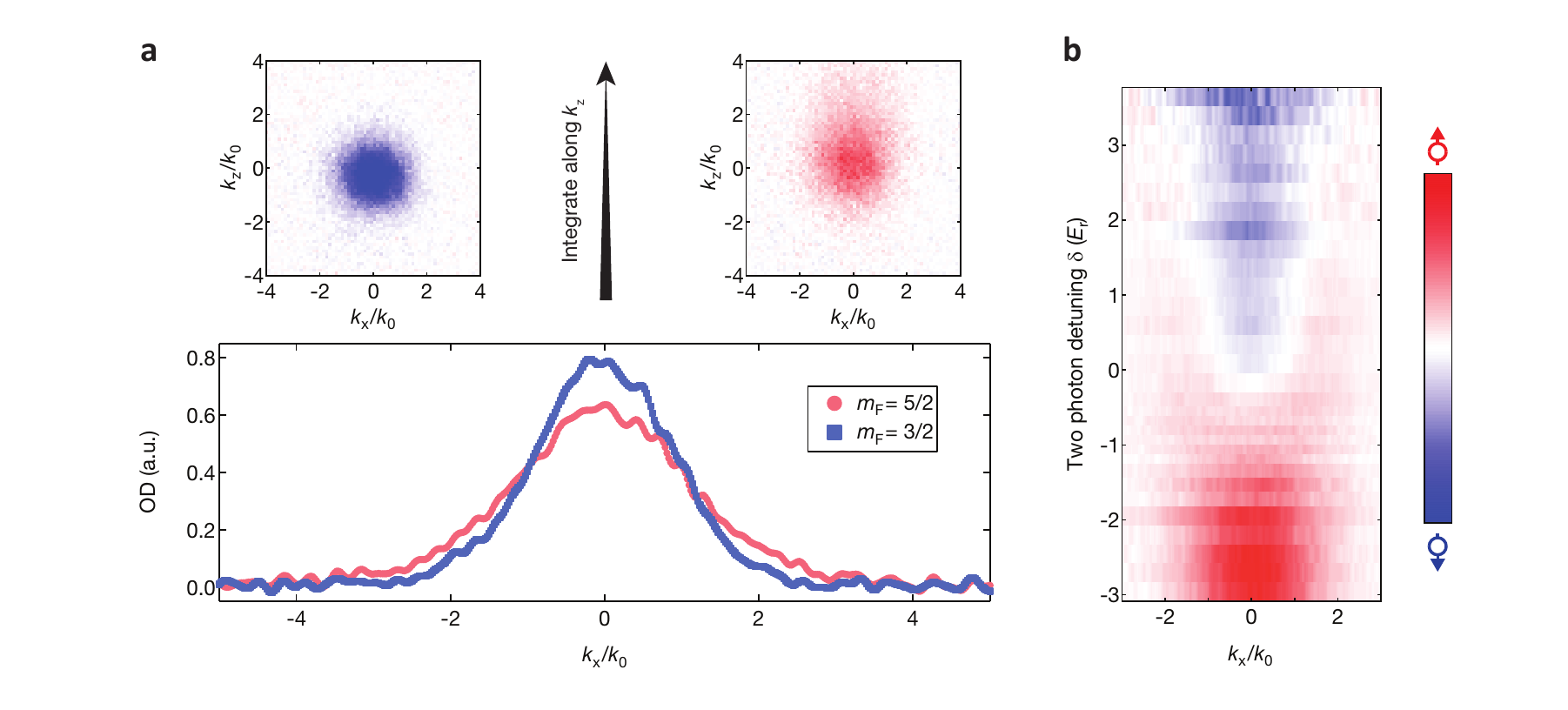}
	\caption*{{\bf Figure S6 } Spin Sensitive TOF Images. {\bf a}, Top figures are the time-of-flight images of  $|\uparrow\rangle$ and $|\downarrow\rangle$ atoms corresponding $m_{F}=\frac{3}{2} $ and $m_{F}=\frac{5}{2}$ states respectively, and their atomic profiles integrated along the $k_z$ direction . {\bf b}, From the the momentum distribution along the $k_x$ direction, we plot the spin-polarization at $k_x$, $(n_{\uparrow} (k_x)-n_{\downarrow}(k_x))/(n_{\uparrow}(k_x)+n_{\downarrow}(k_x))$ as a function of the two-photon detuning $\delta$. In the topological regime, the $|\downarrow\rangle$-dominated regime (blue) coexists together with $|\uparrow\rangle$-dominated regime (red) in the momentum distribution obtained from time-of-flight expansion. The data represent the average of twenty measurements. }
	\label{fig:supp3tof}
\end{figure}

A degenerate equal mixture  of $N_{\uparrow,\downarrow}= 5\times 10^3$ $^{173}$Yb atoms at $ T/T_{F} =$~0.4 is prepared after forced-evaporative cooling in an optical dipole trap~\cite{Song2016a,Song2016b}, where $T_F$ is the Fermi temperature of the trapped fermions with a trapping frequency of $\overline{\omega}=(\omega_x\omega_y\omega_z)^{\frac{1}{3}}$=$2\pi\times$126~Hz. A quantized axis is fixed by the bias magnetic field of 5~G applied along the $\hat{z}$ direction. Before the two-photon Raman transition that couples $|\uparrow\rangle$ and $|\downarrow\rangle$ states is switched on, the intensity of the $\sigma^-$-polarized light (called a lift beam) is linearly ramped up within 3~ms to lift the degeneracy of the ground levels. Simultaneously we compensate the anti-trapping effect of the blue-detuned lights by increasing the depth of the dipole trap. A pair of Raman and lattice beams, with fixed two-photon detuning $\delta$, is then exponentially ramped to the final value with the time constant of $\tau=$7~ms, which adiabatically loads fermions into the optical Raman lattice. Finally the two-photon detuning $\delta$ (or equivalently $m_z$) is changed on demand depending on the experiments as follows: (I) In the measurement of the topological invariant, the fermions are directly loaded into the Raman lattice with the fixed value of $m_z$, by which we determine the $\mathbb{Z}_2$ invariant out of the spin texture in equilibrium; (II) In the adiabatic control between the topological and trivial regimes, we linearly ramp the $ m_{z} $ from $ 0.5E_{\rm r}$ (topological regime) to $ -1Er $ (red trivial regime) within 6 ms and then tune it back to $ 0.5E_{\rm r}$, followed by the detection of the spin texture during the ramp; (III) In the measurement of  quench dynamics , two photon detuning is suddenly switched from the initial ($ \delta = 0.8 E_{\rm r}$ corresponding to the topological phase) to the final value ($ \delta = -2.1 E_{\rm r} $ corresponding to the trivial phase) and vice versa (see Fig.~S5). Finally, blast pulses are applied and absorption images are taken after 6~ms time-of-flight expansion.

\section{Spin sensitive time-of-flight imaging}
To obtain a time-of-flight image sensitive to the spin state, we apply a 556~nm  blast pulses resonant to $ ^{1}$S$_{0} \rightarrow$$^{3}$P$_{1}(F'=7/2)$ transition before the time-of-flight expansion. Using a 399~nm light resonant to  $ ^{1}$S$_{0} \rightarrow$$^{1}$P$_{1} (F'=7/2)$ transition with a blast pulse, three absorption images are obtained providing a time-of-flight atomic distribution $\mathcal{I}_{m_{F}} $ after killing $m_F$-state atoms for 200-400~$\mu$s. Finally, an atomic distribution of $m_F=5/2$ and $m_F=3/2$ atoms is extracted from $ \mathcal{I}_{m_{F}=5/2} - \mathcal{I}_{m_{F}=5/2,3/2} $ and $ \mathcal{I}_{m_{F}=3/2} - \mathcal{I}_{m_{F}=5/2,3/2} $ respectively. We integrate the whole atomic cloud along $k_z$ direction and obtain the atomic profile in the $k_x$ momentum domain as described in Fig.~S6.

\section*{References}
\bigskip

\end{document}